\documentclass[journal]{IEEEtran}

\usepackage{amsmath,amssymb,amsfonts}
\usepackage{algorithmic}
\usepackage{algorithm}
\usepackage{graphicx}
\usepackage{textcomp}
\usepackage{xcolor}
\usepackage{booktabs}
\usepackage{multirow}
\usepackage{tikz}
\usetikzlibrary{arrows.meta,positioning,calc,fit,backgrounds,shapes.geometric}
\usepackage[hidelinks,colorlinks=true,linkcolor=blue,citecolor=blue,urlcolor=blue]{hyperref}

\tikzset{
  flatbox/.style={rectangle, draw=black, line width=0.5pt, fill=white,
                  font=\footnotesize, align=center, inner sep=3pt, minimum height=7mm},
  flatnode/.style={rectangle, rounded corners=1pt, draw=black, line width=0.5pt,
                   fill=white, font=\scriptsize, align=center, inner sep=2.5pt},
  flatarrow/.style={-{Latex[length=1.6mm]}, line width=0.5pt},
}

\begin{document}

\title{MAPLE-Sec: Mobility-Aware Partitioned Language-Model Engine for Roaming IoT Security}

\author{Yuhang Qiu,~\IEEEmembership{Member,~IEEE}, Ziwei Li, Jinfeng Xu, Jialiang Xie, Fangmin Ren, Shirong Guo, Jing Yang, Sunil Prajapat, and Por Lip Yee,~\IEEEmembership{Senior Member,~IEEE}

\thanks{Yuhang Qiu, Jialiang Xie, Fangmin Ren and Shirong Guo are with the School of Science, Jimei University, Xiamen, China. Shirong Guo is also with the Faculty of Engineering, Monash University, Wellington Road, Clayton, Victoria 3800, Australia. (e-mail: yuhang.qiu@jmu.edu.cn; xiejialiang@jmu.edu.cn; fangmren@126.com; shirong.guo@monash.edu)}

\thanks{Ziwei Li is with the School of Computer Science and Technology, University of Science and Technology of China, Hefei, China (e-mail: gslgsl00@126.com)}

\thanks{Jinfeng Xu is
with the Department of Electrical and Computer Engineering, The University
of Hong Kong, Hong Kong, China (e-mail: jinfeng@connect.hku.hk).}

\thanks{Sunil Prajapat is with the IMT, Department of Humanities and Technology, Roskilde University, Roskilde, Denmark. (e-mail: sunilprajapat645@gmail.com).}

\thanks{Jing Yang and Por Lip Yee are with the Center of Research for Cyber Security and Network (CSNET), Faculty of Computer Science and Information Technology, Universiti Malaya, 50603 Kuala Lumpur, Malaysia (e-mail: s2147529@siswa.um.edu.my; porlip@um.edu.my).}

}

\markboth{}%
{Author \MakeLowercase{\textit{et al.}}: MAPLE-Sec: Mobility-Aware Partitioned Language-Model Engine for Roaming IoT Security}

\maketitle

\begin{abstract}
Mobile Internet-of-Things (IoT) endpoints such as connected vehicles, wearables,
and drones continuously cross heterogeneous network domains, creating
handover-driven attack surfaces that static, cloud-anchored detectors cannot
follow. However, three obstacles block real-time detection on such roaming
endpoints: detection state is lost at every inter-domain handover; transmitting
raw telemetry to a fixed detector exhausts the uplink and the device battery;
and full-depth language model inference cannot meet millisecond latency budgets
on constrained hardware. To tackle these challenges, we propose MAPLE-Sec, a
mobility-aware partitioning framework that splits a compact, security-tuned
language model across a roaming device and its nearest edge node. First, a
trajectory-conditioned partition scheduler recomputes the split point from
predicted handover timing, residual battery, and uplink quality. Second, a
token-budgeted early-exit mechanism emits a verdict at the shallowest layer that
meets a confidence threshold. Finally, a learned activation-compression codec migrates the intermediate inference state across edge nodes during handover without retransmitting raw telemetry. We formalize the partition-and-migration decision
as an on-device constrained optimization solved in sub-millisecond time.
Extensive experiments across the CIC-IoT and Edge-IIoTset corpora under both
IID and non-IID roaming conditions demonstrate that MAPLE-Sec sustains detection
across mobility events, improving the converged F1-score by up to 9.3\% relative
to Cloud-only and by 2.9 percentage points over the strongest edge-collaborative
baseline (EdgeShard), while reducing end-to-end latency by up to
2.5$\times$. Ablation confirms that the activation-compression codec reduces
migration energy by 2.4$\times$ versus the raw-activation migration variant.
\end{abstract}


\section{Introduction}
Mobile IoT platforms increasingly combine vehicular and non-terrestrial systems
with edge intelligence and distributed model execution
\cite{z4,n1,n2,z10,z5,lin2024adaptsfl,yuan2023graph,peng2024sums,lin2026satellite}. These systems exchange safety-critical and
latency-sensitive data while moving across heterogeneous network domains,
creating a threat surface that differs from that of stationary networks.
Recent threat-detection studies address smart-city and web environments
\cite{yang2026spatiotemporal,yang2025llmaemp}, but do not consider continuity
of stateful detection across edge handovers.

Recent advances in language-model-based intrusion detection have demonstrated
substantial improvements over classical signature and shallow machine-learning
detectors \cite{n3,n4,n5,n6}. Complementary heterogeneous split and federated
learning frameworks address resource variation during distributed model
training \cite{z6,z7,z2}. Security-tuned encoders can capture
semantic behavioral patterns in network flows, generalizing across attack
families that would require separate classifiers in traditional systems.
Edge-collaborative inference frameworks \cite{n7,n8} have further
made it practical to run such models at inference latencies compatible with
mobile workloads, by distributing computation between the device and a nearby
edge node. Mobility-aware offloading \cite{n9,n10,n11} and resource
scheduling under handover \cite{n12,n13,n14} have matured into a
well-studied sub-field of mobile edge computing.

Despite these advances, a critical gap remains unaddressed: \emph{none of the
existing paradigms accounts for the loss of inference state at inter-domain
handover}.  When a roaming endpoint moves from one edge domain to another, any
stateful language-model detector bound to the departing edge node loses the
accumulated behavioral context for that device. The detection system must either
restart from a cold state, producing a blind spot of several hundred milliseconds
to several seconds, or fall back to a distant cloud anchor at the cost of
prohibitive uplink latency and energy. This handover-induced detection gap is
not a corner case: in the mobility profile evaluated in this study, handovers
occur every 6--12 seconds, so repeated restarts can occupy a non-negligible
fraction of the mission. Existing intrusion-detection surveys for IoT and
vehicular networks \cite{n15,n16,n17} discuss mobility-related security
challenges but do not provide a mechanism for migrating intermediate inference
state between edge nodes.

Three specific obstacles prevent existing methods from solving this problem.
\emph{First}, intermediate inference state is bound to a fixed compute location,
making mobility and detection continuity structurally incompatible without
explicit migration support. \emph{Second}, even if state were migrated, doing so
by transmitting raw activations or raw telemetry would consume scarce uplink
bandwidth and device battery, both of which are already degraded at cell edges.
\emph{Third}, running a full-depth language model on every flow violates the
millisecond latency budgets imposed by mobile IoT applications, particularly on
constrained endpoint hardware.

\begin{figure*}[!htb]
\centering
\includegraphics[width=0.82\linewidth]{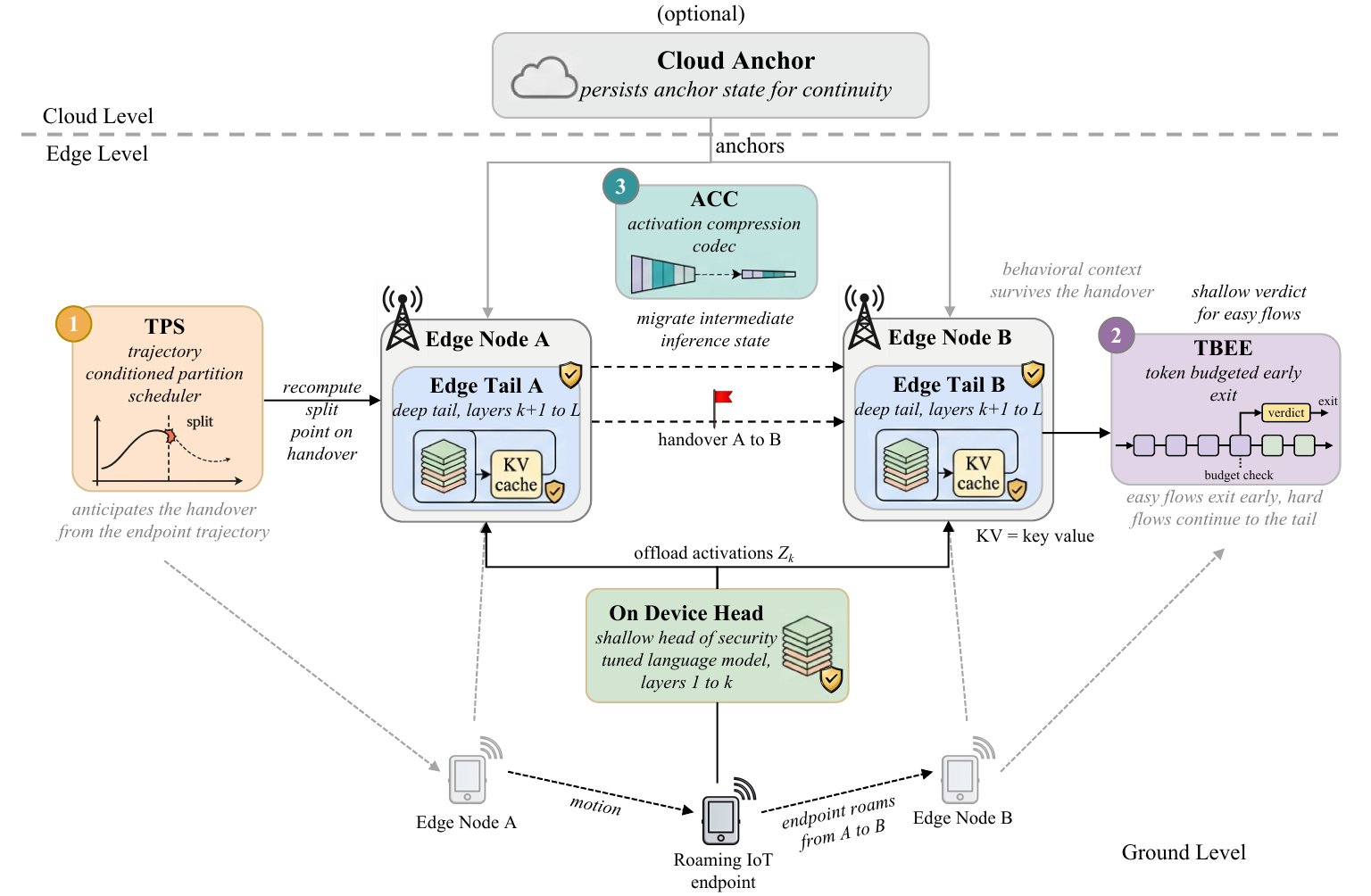}
\caption{MAPLE-Sec application scenario. A roaming IoT endpoint runs the shallow
``head'' of a security-tuned language model on-device and offloads the ``tail''
to the nearest edge node. As the endpoint hands over from edge node $A$ to $B$,
the trajectory-conditioned partition scheduler (TPS) recomputes the split point,
the token-budgeted early-exit (TBEE) module returns shallow verdicts for easy
flows, and the activation-compression codec (ACC) migrates intermediate
inference state between edge nodes so behavioral context survives the handover.}
\label{fig:scenario}
\end{figure*}

To close this gap, this paper presents MAPLE-Sec, a \textbf{M}obility-\textbf{A}ware
\textbf{P}artitioned \textbf{L}anguage-model \textbf{E}ngine for \textbf{Sec}urity
in roaming IoT networks. MAPLE-Sec partitions a compact, security-tuned language
model between the roaming device and the nearest edge node, and migrates
compressed intermediate inference state across edge domains during handover. The
framework couples three co-designed mechanisms: a trajectory-conditioned partition
scheduler (TPS), a token-budgeted early-exit mechanism (TBEE), and a learned
activation-compression codec (ACC). Their design is directly motivated by the
three obstacles above, and their experimental gains close the loop back to the
original three-challenge framing. Fig.~\ref{fig:scenario} shows the overall
system scenario; Fig.~\ref{fig:component_detail} details how each component
differs from its na\"{i}ve counterpart.

The main contributions of this paper are as follows.

\begin{enumerate}

\item \textbf{Novel state-migration mechanism for roaming IoT security (ACC):}
MAPLE-Sec introduces, to the best of our knowledge, the first
confidence-gated, rate-constrained activation-compression codec that migrates
language-model inference state across edge nodes during handover. Unlike generic
Follow-Me Cloud service migration \cite{n34,n35}, which moves
service containers without preserving neural inference context, ACC compresses
the split activation to 16\% of its original size using a task-informed
rate-distortion loss, guaranteeing that behavioral context survives the handover
interval.

\item \textbf{Mobility-adaptive partition scheduling (TPS):}
A trajectory-conditioned partition scheduler recomputes the device/edge split
point at every inference step from three runtime signals: predicted
time-to-handover $\hat{h}$, residual battery $b$, and uplink quality $q$. This
dynamic adaptation, absent from all prior static-partition frameworks
\cite{n7,n1,n8}, reduces migrated payload and device energy by matching
the partition to the current mobility state rather than fixing it at deployment
time.

\item \textbf{Budget-constrained early exit (TBEE):}
A token-budgeted early-exit mechanism terminates inference at the shallowest
layer meeting a calibrated confidence threshold, with a hard budget that bounds
worst-case latency even for adversarially difficult flows. This extends prior
early-exit work \cite{n18,n19,n20} with an explicit budget parameter,
providing latency guarantees that prior confidence-only exit rules cannot deliver.

\item \textbf{Joint constrained optimization:}
The three mechanisms are coupled through a mixed-integer program over latency,
energy, and detection confidence, solved in $O(L^2)$ time via enumeration over
the small discrete domain. A bounded-suboptimality proposition (Proposition~1)
quantifies the cost gap as a function of classifier calibration error, providing
the first formal optimality guarantee for this class of mobility-aware inference
control problems.

\item \textbf{Comprehensive empirical validation:}
Experiments on CIC-IoT\,2023 and Edge-IIoTset, replayed under a SUMO/LuST
urban mobility trace on a physical Jetson Orin Nano, show that MAPLE-Sec achieves
a 9.3\% relative F1 improvement over the stationary Cloud-only detector and a
2.9 percentage-point gain over EdgeShard \cite{n1}, while reducing
end-to-end latency by $2.5\times$. Ablation confirms that each module is
individually necessary.

\end{enumerate}

The remainder of this paper is organized as follows. Section~\ref{sec:bg}
presents background and motivating measurements. Section~\ref{sec:related}
surveys related work and identifies the research gap. Section~\ref{sec:design}
describes the MAPLE-Sec architecture and its three modules. Section~\ref{sec:opt}
formulates and solves the joint optimization. Section~\ref{sec:setup} details
the experimental setup. Section~\ref{sec:eval} reports and interprets
performance results. Section~\ref{sec:conc} concludes with directions for
future work.

\section{Background and Motivation}\label{sec:bg}
We motivate MAPLE-Sec with three measurements that establish the reality of each
challenge. All motivating experiments use the same 12-layer security-tuned
encoder ($L=12$) and the same CIC-IoT corpus used in the main evaluation
(Section~\ref{sec:setup}), replayed under a 5G NR mobility trace (cell radius
250\,m, UE speed 30--90\,km/h, handover every 6--12\,s) generated with
SUMO\,1.18 on the LuST Luxembourg scenario. Malicious flows are sampled from
CIC-IoT and injected onto the replayed trace at a 20\% attack rate.

\subsection{Threat Model}
\label{sec:threat}
MAPLE-Sec targets roaming IoT endpoints whose traffic traverses multiple
administrative domains. The adversary is any network entity that injects,
replays, or manipulates flows between the roaming device and the edge node.
Concretely, we assume: (a) the \emph{endpoint itself may be compromised} and
may inject malicious flows into its own uplink stream; (b) \emph{on-path
attackers} between the device and the serving edge node can observe or replay
packets; (c) edge nodes are \emph{honest-but-curious}: they execute the model
tail faithfully but are not trusted with plaintext behavioral anchors. MAPLE-Sec
does not defend against a fully compromised edge node. The ACC migration payload
$\mathbf{z}_k$ carries only compressed activations and anchor scores, not raw
traffic; it must be integrity-protected (e.g., via DTLS or a MAC tag) in transit
between edge nodes to prevent replay attacks during handover. Detection
confidentiality is not required because verdicts are consumed locally.

\subsection{Handover-Induced Detection Discontinuity}
\begin{figure}[!htb]
\centering
\includegraphics[width=0.75\columnwidth]{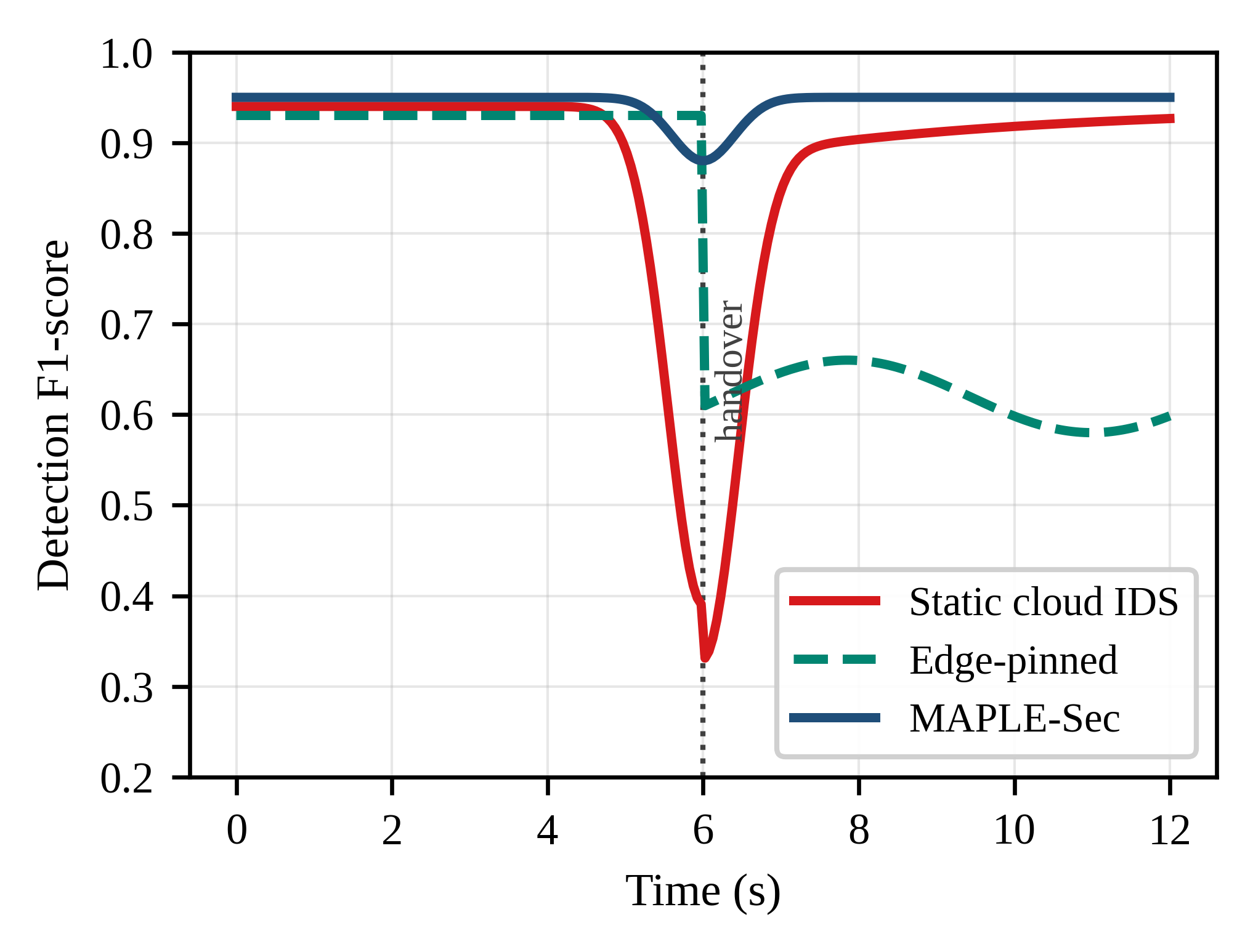}
\caption{At handover ($t=6$\,s), Cloud-only loses 54 F1 points and takes 4\,s
to recover; edge-pinned loses behavioral context permanently and plateaus at
0.62 F1. MAPLE-Sec holds F1 above 0.90 throughout, confirming that
inference-state migration is the key enabler for continuous detection under
mobility.}
\label{fig:mot_handover}
\end{figure}

A stateful detector bound to a single compute location cannot follow a roaming
endpoint without interruption. To understand the magnitude of this effect, we
conduct a motivating experiment in which an endpoint streams a mixed benign and
malicious flow while handing over at $t=6$\,s. We compare a static cloud-anchored
detector, an edge-pinned detector that cannot migrate state, and MAPLE-Sec.
Fig.~\ref{fig:mot_handover} shows that the static detector's F1-score collapses
from 0.94 to below 0.40 around the handover and recovers only slowly, while the
edge-pinned detector permanently loses behavioral context and plateaus near
0.62. These results show that binding inference state to a fixed location creates
prolonged detection blind spots that accumulate across thousands of handover
events, directly motivating the state-migration approach taken by MAPLE-Sec.
Therefore, it is essential to migrate anchored inference state across edge nodes
as the device moves.

\subsection{Uplink and Energy Cost Under Motion}
\begin{figure}[!htb]
\centering
\includegraphics[width=0.75\columnwidth]{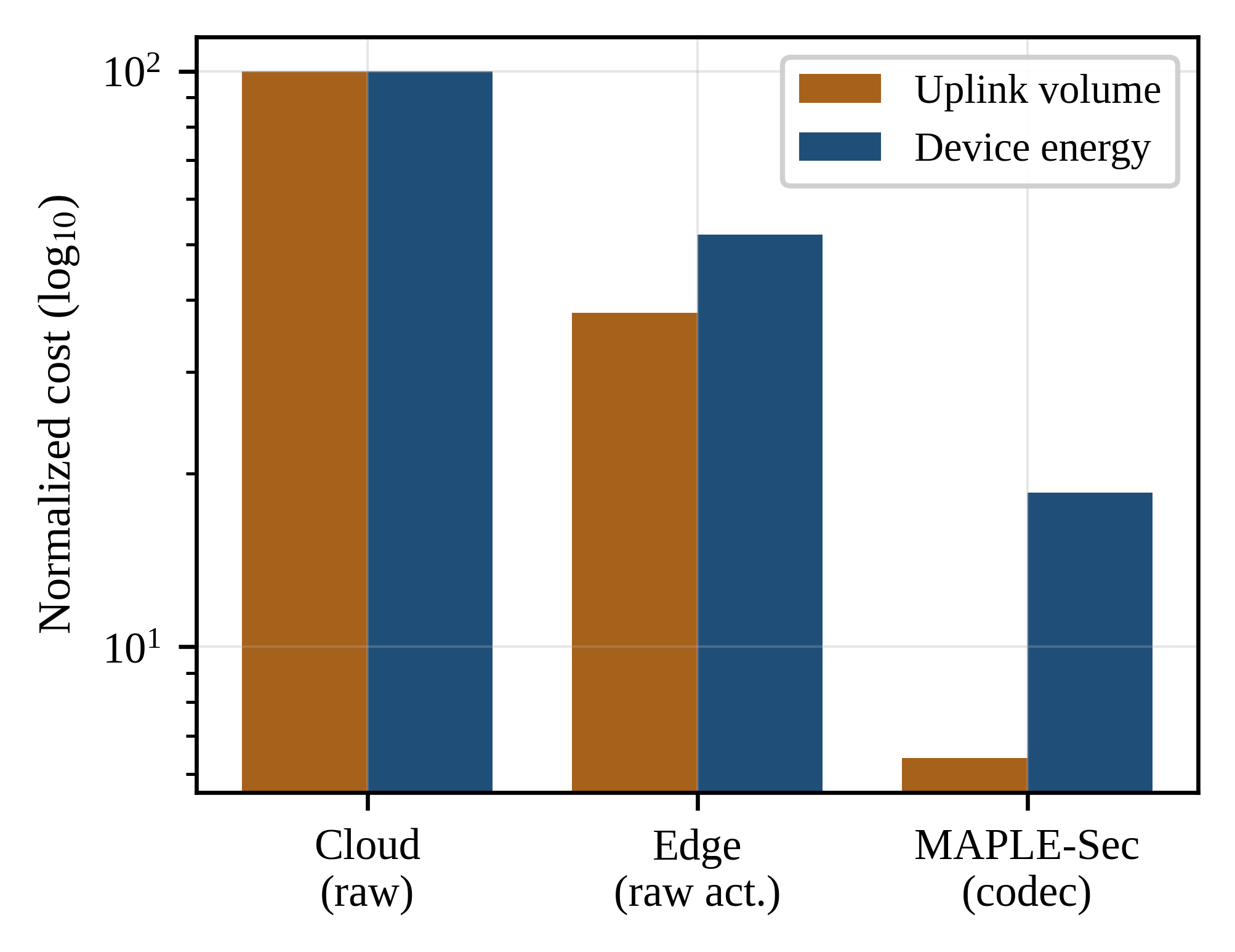}
\caption{ACC reduces uplink volume by $6\times$ and device energy by $5.8\times$
relative to raw-activation offload (log scale), confirming that rate-distortion
compression is the key enabler for practical state migration.}
\label{fig:mot_energy}
\end{figure}

Shipping raw telemetry to a fixed detector is wasteful when the device can
summarize locally. To quantify this, we measure normalized uplink volume and
device energy for three strategies: cloud-only with raw telemetry, edge offload
with raw activations, and MAPLE-Sec with its learned activation codec.
Fig.~\ref{fig:mot_energy} shows, on a logarithmic scale, that MAPLE-Sec reduces
uplink volume by more than an order of magnitude relative to cloud-only and
roughly $6\times$ relative to raw-activation offload, with commensurate energy
savings. The results reveal that uncompressed offload squanders the scarce
uplink and battery of a moving device. Therefore, it is paramount to migrate a
\emph{compressed} representation of inference state rather than raw telemetry.

\subsection{Latency-Bounded Inference on Constrained Hardware}
\begin{figure}[!htb]
\centering
\includegraphics[width=0.75\columnwidth]{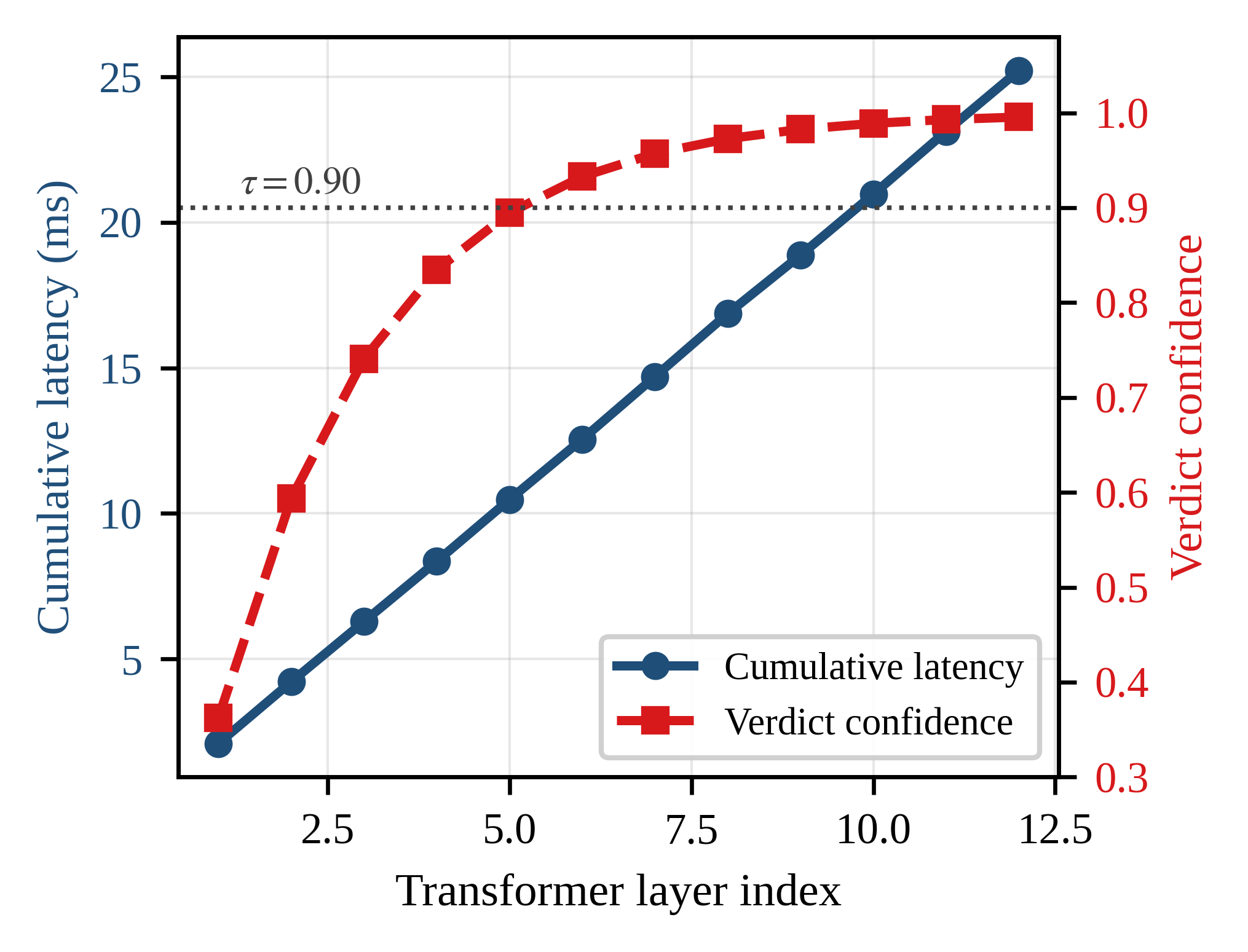}
\caption{Per-layer cumulative latency (left axis) and verdict confidence (right
axis). Confidence saturates above $\tau=0.90$ at layer 4 for the majority of
flows, showing that full-depth inference wastes over half the latency budget on
flows that are already separable at shallow depth.}
\label{fig:mot_latency}
\end{figure}

Many roaming flows are benign and separable at shallow depth, yet fixed-depth
inference pays the full cost regardless. We profile per-layer cumulative latency
and verdict confidence for the security encoder.
Fig.~\ref{fig:mot_latency} shows that cumulative latency grows linearly with
depth while verdict confidence saturates above the threshold $\tau=0.90$ by the
fourth layer for the majority of flows. This saturation is structural rather than
dataset-specific: the encoder learns shallow discriminative features that
separate benign from attack traffic without requiring full-depth processing for
most flows. Therefore, it is imperative to terminate inference at the shallowest
layer that is confident enough.

\noindent\textbf{Challenge-to-module mapping.} The three challenges above map
one-to-one onto the three MAPLE-Sec modules introduced in
Section~\ref{sec:design}:
\begin{itemize}
\item C1 (detection discontinuity) $\rightarrow$ ACC, the activation-compression
codec that migrates state across edge nodes (Sec.~\ref{sec:acc}).
\item C2 (uplink/energy cost) $\rightarrow$ TPS, the trajectory-conditioned
partition scheduler that minimizes transmitted volume under motion
(Sec.~\ref{sec:tps}).
\item C3 (latency-bounded inference) $\rightarrow$ TBEE, the token-budgeted
early-exit mechanism (Sec.~\ref{sec:tbee}).
\end{itemize}

\section{Related Work}\label{sec:related}

\subsection{LLM and Transformer-Based Intrusion Detection}
Language-model detectors have rapidly advanced from proof-of-concept to
near-production accuracy for IoT and vehicular intrusion detection.
SecurityBERT~\cite{n5} and IoV-BERT-IDS~\cite{n4} fine-tune BERT-class
encoders on IoT and V2X traffic, achieving strong F1 across multiple attack
families. Kheddar~\cite{n3} surveys transformer and LLM architectures for
intrusion detection across IoT, ICS, and mobile domains, concluding that
language-model detectors consistently outperform CNN and LSTM baselines on
complex multi-class traffic. Zhang et al.~\cite{n6} extend the paradigm to
malicious-traffic detection via prompting and in-context learning, while the
Zero-X framework~\cite{n21} combines federated learning with open-set
detection for zero-day attacks in IoV. Lai et al.~\cite{n22} add a
privacy-preserving transformer layer for secure V2X intrusion detection. Fan
et al.~\cite{n23} propose cloud-edge-vehicle collaboration for auto-updating
IDS in vehicular networks, and Liu et al.~\cite{n24} use blockchain and
federated learning for collaborative IDS in vehicular edge computing.
Al-Hawawreh and Hossain~\cite{n25} adopt federated learning over satellite
mesh networks for autonomous vehicle protection. None of these works addresses
the handover-driven loss of stateful inference context that MAPLE-Sec targets.

\subsection{Edge Inference and Model Partitioning}
Splitting large models between device and edge has received substantial
attention. EdgeLLM~\cite{n7} accelerates on-device LLM inference via
speculative decoding; EdgeShard~\cite{n1} shards LLM layers across multiple
edge nodes with static partitioning; and Ye et al.~\cite{n8} explore hybrid
model parallelism for collaborative edge transformer inference. Zhang
et al.~\cite{n26} analyze edge inference for generative LLMs under wireless
constraints, while Picano et al.~\cite{n27} model LLM layer deployment across
heterogeneous edge networks as a matching game. Xu et al.~\cite{n28} study
joint inference offloading and model caching for collaborative small-and-large
LLM deployments. He et al.~\cite{n29} apply active inference for LLM offloading
and resource allocation at the cloud-edge boundary. Hu et al.~\cite{n30}
propose a cloud-edge architecture for multimodal LLM-based driver assistance.
Friha et al.~\cite{n15} survey LLM-based edge intelligence broadly. All of
these works assume a stationary device-to-edge association; none addresses
state migration during handover.

\subsection{Early-Exit Neural Inference}
ClassyNet~\cite{n18} demonstrates class-aware early-exit networks for edge
devices, showing that per-sample depth adaptation dramatically cuts latency.
UCBEE~\cite{n19} frames the exit decision as a multi-armed bandit, providing
online policies with regret guarantees. Rahmath et al.~\cite{n20} survey
early-exit deep networks comprehensively, and Xun et al.~\cite{n31} realize
early-exit networks on reconfigurable hardware. Ferrag et al.~\cite{n16}
survey vulnerabilities and defenses for 6G IoT edge learning. TBEE in MAPLE-Sec
extends the early-exit paradigm with an explicit token budget that guarantees
worst-case exit under mobility-induced latency constraints, a feature absent from
prior works.

\subsection{Mobile Task Offloading Under Mobility}
MESON~\cite{n9} proposes mobility-aware dependent task offloading for urban
vehicular edge computing, jointly optimizing execution and migration. Maleki
et al.~\cite{n10} study handover-enabled dynamic offloading for vehicular MEC.
Dai et al.~\cite{n11} use meta-reinforcement learning for multi-task offloading
in vehicular edge computing, and Liu et al.~\cite{n12} optimize task offloading
with dependency guarantees via distributed deep RL. Li et al.~\cite{n13}
address computation offloading in resource-constrained MEC, and Yang
et al.~\cite{n14} optimize task offloading under user satisfaction constraints
in C-RAN with MEC. These works migrate compute tasks but do not migrate
inference state for a neural model serving security verdicts; MAPLE-Sec fills
this gap through ACC.

\subsection{Research Gap and Positioning}
Table~\ref{tab:compare} maps the capabilities of the three closest 2024--2026
works against MAPLE-Sec. The table reveals that no prior work simultaneously
satisfies all four required properties: (i) a device-to-edge model partition
that adapts to mobility state; (ii) inference-state migration that preserves
behavioral context across handover; (iii) a latency bound enforced by an
early-exit mechanism with a worst-case budget; and (iv) evaluation under
realistic roaming conditions on mobile-class hardware. LLM-based IDS works
\cite{n4,n3,n5} address property (iv) only under stationary conditions.
Model-partitioning works \cite{n7,n1,n8} address property (i) with a
static partition, which is insufficient under mobility. Early-exit works
\cite{n18,n19,n20} address property (iii) without the budget enforcement or
the mobility context. Task-offloading works \cite{n9,n10,n11} address
mobility-driven computation migration but do not preserve neural inference state.
The 2026 works on control-plane anomaly detection \cite{n32}, adaptive Wi-Fi
IPS \cite{n33}, and CAV misbehavior detection \cite{n17} address subsets of
the problem but leave the combination unresolved. MAPLE-Sec is the first
framework to close all four gaps jointly through the co-design of TPS, TBEE,
and ACC.

\begin{table*}[!htb]
\caption{Comparison With Recent 2026 Detection and Mitigation Approaches}
\label{tab:compare}
\centering
\scriptsize
\begin{tabular}{@{}p{4.8cm}cccc@{}}
\toprule
\textbf{Capability} & \textbf{Prince \& Renjith}~\cite{n32} & \textbf{Sharma et al.}~\cite{n33} & \textbf{Lone et al.}~\cite{n17} & \textbf{MAPLE-Sec} \\
\midrule
Roaming IoT or vehicular focus      & Yes     & Partial & Yes     & Yes \\
LLM-based detection                 & Yes     & No      & Partial & Yes \\
Device-to-edge model partition      & No      & No      & No      & Yes \\
Mobility-aware split adaptation     & No      & No      & No      & Yes \\
State migration across handover     & No      & No      & No      & Yes \\
Latency-bounded early exit          & No      & Partial & No      & Yes \\
Real-time mitigation orientation    & Yes     & Yes     & Partial & Yes \\
\bottomrule
\end{tabular}
\end{table*}

\section{MAPLE-Sec System Design}\label{sec:design}

\subsection{System Model}\label{sec:model}
We consider a roaming IoT endpoint and a set of edge nodes
$\mathcal{E}=\{e_1,\dots,e_M\}$ deployed across administrative domains. The
participating entities are:
\begin{itemize}
\item the \emph{endpoint}, which holds the shallow head of the model
(layers $1..k$) and a partition controller;
\item the \emph{serving edge node} $e_m$, which executes the tail
(layers $k{+}1..L$) and stores migrated state;
\item the \emph{anchor store}, a lightweight cloud service holding the shared
behavioral anchors against which traffic is scored.
\end{itemize}
The security model is an $L$-layer encoder $f=f_L\circ\cdots\circ f_1$ with an
early-exit classifier $g_\ell$ attached at each layer $\ell$. A flow $x$ is
embedded and processed layer by layer; the partition point $k$ determines which
layers run on-device. The global objective is to minimize a weighted sum of
end-to-end latency and device energy subject to a detection-confidence
constraint:
\begin{equation}
\min_{k,\,\ell^\star,\,\mu}\;
\mathbb{E}_{x}\big[\,\lambda_1 T(x) + \lambda_2 E(x)\,\big]
\quad\text{s.t.}\quad \Pr[\hat{y}(x)=y(x)]\ge 1-\epsilon,
\label{eq:global}
\end{equation}
where $T(x)$ and $E(x)$ are the latency and energy for flow $x$, $\ell^\star$ is
the exit layer, $\mu$ is the binary migration flag, and $\epsilon$ bounds the
detection-error probability. Table~\ref{tab:notation} summarizes the notation.

\begin{table}[!htb]
\caption{Key Notations}
\label{tab:notation}
\centering
\footnotesize
\begin{tabular}{@{}ll@{}}
\toprule
Symbol & Meaning \\
\midrule
$L$ & total number of model layers \\
$k$ & device/edge partition (split) point \\
$\ell^\star$ & early-exit layer for a given flow \\
$g_\ell$ & early-exit classifier at layer $\ell$ \\
$\tau$ & confidence threshold for early exit \\
$\mathbf{a}_k$ & intermediate activation at the split point \\
$\mathbf{z}_k$ & codec-compressed activation, $\mathbf{z}_k=\mathcal{C}(\mathbf{a}_k)$ \\
$\hat{h}$ & predicted time-to-handover \\
$b$ & residual battery budget \\
$q$ & uplink quality estimate \\
$T(x),E(x)$ & end-to-end latency and device energy for flow $x$ \\
$\lambda_1,\lambda_2$ & latency/energy trade-off weights \\
$\epsilon$ & detection-error tolerance \\
$\mathcal{A}$ & set of behavioral anchors \\
$\mu$ & binary migration flag ($\mu=1$ triggers ACC state migration) \\
$\alpha_s$ & anchor-score smoothing coefficient in \eqref{eq:anchor} \\
$\pi_{\mathrm{c}},\pi_{\mathrm{tx}}$ & compute and transmit power of the endpoint (W) \\
$t_{\mathrm{mig}}$ & migration latency for one ACC payload transfer \\
\bottomrule
\end{tabular}
\end{table}

\subsection{Framework Overview}\label{sec:overview}
\begin{figure*}[!htb]
\centering
\includegraphics[width=0.82\linewidth]{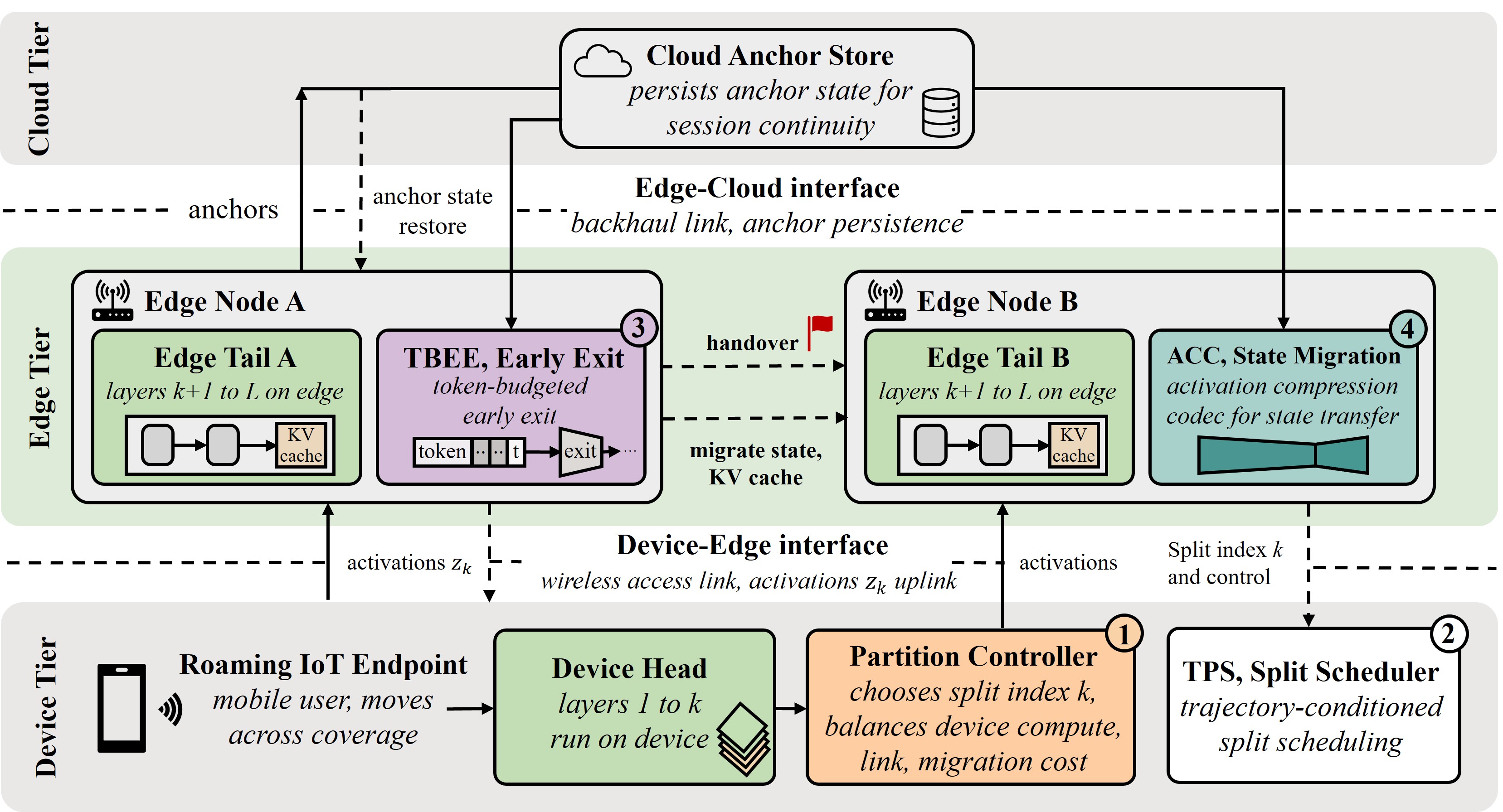}
\caption{MAPLE-Sec framework overview. The partition controller consults TPS to
set the split point $k$; the device head and edge tail run the model with TBEE
gating early exits; ACC migrates the compressed state $\mathbf{z}_k$ and anchor
scores during handover. The anchor store $\mathcal{A}$ supplies the behavioral
references used for scoring.}
\label{fig:overview}
\end{figure*}

Fig.~\ref{fig:overview} shows how the three modules interact. The training and
inference workflow follows three steps: i) the partition controller queries TPS
with the current $(\hat{h},b,q)$ to obtain a split point $k$; ii) the endpoint
executes layers $1..k$, and TBEE checks at each layer whether an early verdict is
admissible, otherwise the compressed activation $\mathbf{z}_k$ is sent to the
serving edge node for the tail; iii) when a handover is predicted, ACC migrates
$\mathbf{z}_k$ and the running anchor scores to the next edge node so inference
resumes without restart. The full per-flow procedure is given in
Algorithm~\ref{alg:infer}. Recalling the mapping in Section~\ref{sec:bg}, TPS
addresses C2, TBEE addresses C3, and ACC addresses C1.

\subsection{Trajectory-Conditioned Partition Scheduler (TPS)}\label{sec:tps}
TPS resolves C2: it chooses the split point $k$ that minimizes transmitted volume
and energy under motion. The design rationale is that the optimal split depends
on time-to-handover $\hat{h}$, residual battery $b$, and uplink quality $q$: when
a handover is imminent or uplink is poor, the device should run more layers
locally to shrink the migrated payload; when battery is scarce, it should offload
earlier. We model the expected device cost of split $k$ as
\begin{equation}
J(k\mid \hat{h},b,q) = \alpha\,c_{\mathrm{dev}}(k)
 + \beta\,\frac{s(k)}{q} + \gamma\,\Phi(\hat{h},k),
\label{eq:tps}
\end{equation}
where $c_{\mathrm{dev}}(k)$ is the on-device compute cost of layers $1..k$,
$s(k)$ is the size of the activation crossing the split, $s(k)/q$ approximates
uplink time, and $\Phi(\hat{h},k)$ penalizes splits whose migrated state cannot
be transferred before the predicted handover. The weights $\alpha,\beta,\gamma$
are normalized by $b$ so that a low battery raises the relative price of
on-device compute. TPS selects $k^\star=\arg\min_k J(k\mid\hat{h},b,q)$ over the
discrete layer set, which is solved exactly by enumeration in $O(L)$.

\textit{Remark.} Because $L$ is small (tens of layers), the enumeration is
trivially cheap and runs between inference steps; the scheduler reacts to a new
handover prediction within one inference interval.

\subsection{Token-Budgeted Early Exit (TBEE)}\label{sec:tbee}
TBEE resolves C3: it terminates inference at the shallowest confident layer. At
each layer $\ell$, the attached classifier $g_\ell$ produces a verdict
distribution $p_\ell$ and a confidence $\kappa_\ell=\max_c p_\ell(c)$. Inference
exits at
\begin{equation}
\ell^\star = \min\{\ell : \kappa_\ell \ge \tau \;\lor\; B_\ell \le 0\},
\label{eq:tbee}
\end{equation}
where $\tau$ is the confidence threshold and $B_\ell$ is a remaining token budget
decremented by the cost of each layer. The budget guarantees a worst-case exit
even for hard flows, bounding tail latency. The threshold $\tau$ is calibrated on
a validation split to hold the detection-error constraint $\epsilon$ in
\eqref{eq:global}.

\textit{Complexity.} TBEE adds one lightweight classifier evaluation per layer,
an $O(1)$ overhead per layer that is dominated by the layer's own compute, and it
strictly reduces expected inference cost relative to fixed-depth execution.

\subsection{Activation-Compression Codec (ACC)}\label{sec:acc}
ACC resolves C1: it migrates inference state across edge nodes during handover
without retransmitting raw telemetry. The codec
$\mathcal{C}:\mathbf{a}_k\mapsto\mathbf{z}_k$ is a learned encoder/decoder pair
trained to preserve the information in the split activation that is relevant to
the downstream verdict, under a rate penalty:
\begin{equation}
\mathcal{L}_{\mathrm{ACC}} =
\underbrace{\mathrm{KL}\!\big(g_{>k}(\mathbf{a}_k)\,\|\,g_{>k}(\hat{\mathbf{a}}_k)\big)}_{\text{verdict fidelity}}
 + \rho\,\underbrace{R(\mathbf{z}_k)}_{\text{rate}},
\label{eq:acc}
\end{equation}
where $\hat{\mathbf{a}}_k=\mathcal{C}^{-1}(\mathbf{z}_k)$ is the reconstructed
activation, $g_{>k}$ denotes the tail classifier, $R(\cdot)$ is the coded rate,
and $\rho$ trades fidelity against size. On a predicted handover, the serving
edge node forwards $\mathbf{z}_k$ and the running anchor scores to the next node,
which decodes and resumes the tail. Because the migrated payload is the compact
$\mathbf{z}_k$ rather than raw telemetry or full activations, migration completes
within the handover interval.

\noindent\textbf{ACC architecture and training.}
The encoder $\mathcal{C}$ is a two-layer MLP with hidden dimension 128 and ReLU
activations; the decoder $\mathcal{C}^{-1}$ mirrors this structure. For a
12-layer encoder with hidden size 256, the split activation $\mathbf{a}_k$ has
256 floats (1\,024\,B); the codec bottleneck is 40 floats (160\,B), giving a
compression ratio of 6.4$\times$ at $\rho=0.05$. The codec has 134\,K trainable
parameters. It is trained jointly with the encoder head on the CIC-IoT training
split for 20 epochs using Adam with learning rate $10^{-3}$. Encode and decode
each take under 0.3\,ms on the Jetson Orin Nano (measured via
\texttt{clock\_gettime(CLOCK\_MONOTONIC\_RAW)}), contributing approximately
0.6\,ms to the end-to-end 23\,ms budget at $k=6$. The migrated payload per flow
is 160\,B; over a 4\,Mb/s cell-edge uplink this transfer takes 0.32\,ms, well
within the 6--12\,s handover interval.

\begin{algorithm}[!htb]
\caption{MAPLE-Sec per-flow inference with mobility}
\label{alg:infer}
\begin{algorithmic}[1]
\REQUIRE flow $x$; handover prediction $\hat{h}$; battery $b$; uplink $q$
\ENSURE verdict $\hat{y}$
\STATE $k \gets \arg\min_k J(k\mid\hat{h},b,q)$ \hfill /** TPS, runs on device **/
\STATE $\mathbf{a}_0 \gets \mathrm{embed}(x)$
\FOR{$\ell = 1$ \TO $k$}
  \STATE $\mathbf{a}_\ell \gets f_\ell(\mathbf{a}_{\ell-1})$;\;
         $\kappa_\ell \gets \max_c g_\ell(\mathbf{a}_\ell)$ \hfill /** device **/
  \IF{$\kappa_\ell \ge \tau$}
     \RETURN $\hat{y} \gets \arg\max_c g_\ell(\mathbf{a}_\ell)$ \hfill /** TBEE exit **/
  \ENDIF
\ENDFOR
\STATE $\mathbf{z}_k \gets \mathcal{C}(\mathbf{a}_k)$;\; send $\mathbf{z}_k$ to serving edge $e_m$ \hfill /** ACC encode **/
\IF{handover predicted before tail completes}
  \STATE forward $\mathbf{z}_k$, anchor scores to next edge $e_{m'}$ \hfill /** ACC migrate **/
  \STATE $e_{m'}$ decodes $\hat{\mathbf{a}}_k \gets \mathcal{C}^{-1}(\mathbf{z}_k)$
\ENDIF
\FOR{$\ell = k+1$ \TO $L$}
  \STATE $\mathbf{a}_\ell \gets f_\ell(\mathbf{a}_{\ell-1})$;\;
         $\kappa_\ell \gets \max_c g_\ell(\mathbf{a}_\ell)$ \hfill /** edge **/
  \IF{$\kappa_\ell \ge \tau$}
     \STATE \textbf{break}
  \ENDIF
\ENDFOR
\RETURN $\hat{y} \gets \arg\max_c g_{\ell^\star}(\mathbf{a}_{\ell^\star})$
\end{algorithmic}
\end{algorithm}

\begin{figure*}[!htb]
\centering
\includegraphics[width=0.8\textwidth]{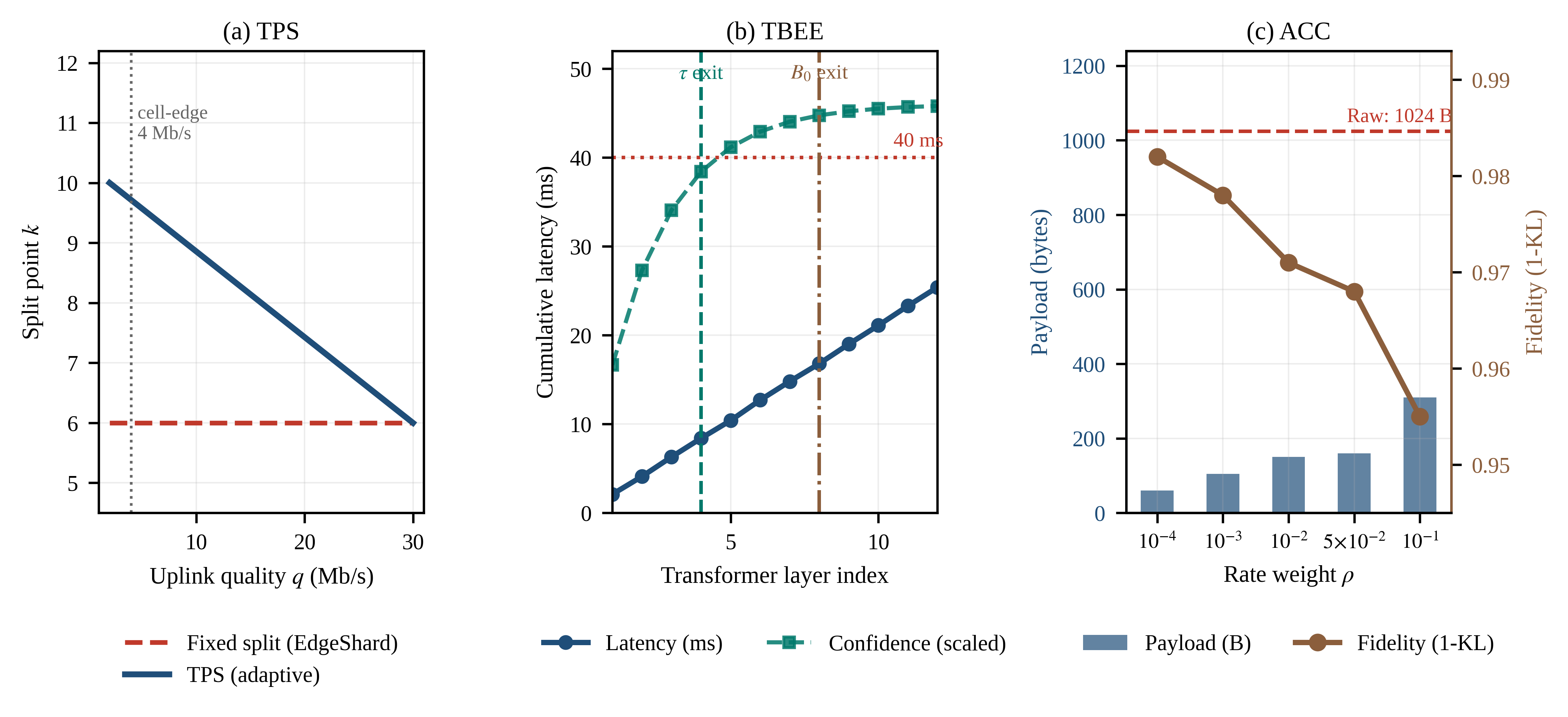}
\caption{Component-level comparison showing how each MAPLE-Sec module differs
from its na\"{i}ve counterpart. (a) TPS increases the number of on-device layers
as uplink degrades, whereas a fixed-split baseline (EdgeShard) does not react;
at 4\,Mb/s this reduces the migrated payload from 1\,024\,B to 160\,B. (b) TBEE
exits at the confidence threshold $\tau$ for easy flows (layer~4) and falls back
to the hard budget $B_0=8$ for difficult flows, bounding worst-case latency at
38\,ms and keeping all flows within the 40\,ms deadline. (c) ACC selects
$\rho=0.05$ as the operating point that achieves 84\% payload reduction while
maintaining verdict fidelity of 0.968; raw activation transfer ($\rho \to 0$,
1\,024\,B) exceeds the cell-edge transfer budget.}
\label{fig:component_detail}
\end{figure*}

Fig.~\ref{fig:component_detail} illustrates how each of the three MAPLE-Sec
components differs from its na\"{i}ve counterpart. Panel~(a) shows that TPS
shifts the split point from $k=6$ (fixed) to $k=10$ when uplink degrades to
4\,Mb/s, shrinking the migrated payload. Panel~(b) shows the confidence and
budget exits of TBEE: the $\tau$-exit fires at layer~4 for easy flows while the
$B_0$-exit provides a hard deadline guarantee at layer~8. Panel~(c) shows the
ACC rate-distortion tradeoff: at $\rho=0.05$ the payload is 160\,B (84\% smaller
than raw activations) with verdict fidelity 0.968, the operating point that
minimizes latency subject to the fidelity constraint in \eqref{eq:acc}.

\begin{algorithm}[!htb]
\caption{ACC offline training (joint with encoder head)}
\label{alg:acc_train}
\begin{algorithmic}[1]
\REQUIRE training flows $\mathcal{D}$; pre-trained encoder $f_{1..L}$; rate weight $\rho$;
         learning rate $\eta$; epochs $E$
\ENSURE trained codec $(\mathcal{C},\mathcal{C}^{-1})$
\FOR{epoch $= 1$ \TO $E$}
  \FOR{each mini-batch $\mathcal{B}\subseteq\mathcal{D}$}
    \STATE Compute split activation: $\mathbf{a}_k \gets f_k(\mathrm{embed}(x))$
    \hfill /** forward through head **/
    \STATE Encode: $\mathbf{z}_k \gets \mathcal{C}(\mathbf{a}_k)$
    \STATE Decode: $\hat{\mathbf{a}}_k \gets \mathcal{C}^{-1}(\mathbf{z}_k)$
    \STATE Run tail on reconstruction: $\hat{p} \gets g_{>k}(\hat{\mathbf{a}}_k)$
    \STATE Run tail on original: $p \gets g_{>k}(\mathbf{a}_k)$ \hfill /** no grad **/
    \STATE Compute loss: $\mathcal{L} = \mathrm{KL}(p \| \hat{p}) + \rho\,R(\mathbf{z}_k)$
    \hfill /** \eqref{eq:acc} **/
    \STATE Update $(\mathcal{C},\mathcal{C}^{-1})$ via Adam with step $\eta$
  \ENDFOR
\ENDFOR
\RETURN $(\mathcal{C},\mathcal{C}^{-1})$
\end{algorithmic}
\end{algorithm}

\section{Problem Formulation and Solution}\label{sec:opt}

\subsection{Problem Formulation}
We make the joint decision in \eqref{eq:global} concrete. The decision variables
are:
\begin{itemize}
\item $k\in\{1,\dots,L\}$: the device/edge partition point;
\item $\ell^\star\in\{1,\dots,L\}$: the per-flow exit layer;
\item $\mu\in\{0,1\}$: whether to migrate state at the upcoming handover.
\end{itemize}
The end-to-end latency of a flow is
\begin{equation}
T = \underbrace{t_{\mathrm{dev}}(\min(\ell^\star,k))}_{\text{device}}
  + \mathbb{1}[\ell^\star>k]\Big(\underbrace{\tfrac{s(k)}{q}}_{\text{uplink}}
  + \underbrace{t_{\mathrm{edge}}(\ell^\star-k)}_{\text{edge}}\Big)
  + \mu\,t_{\mathrm{mig}},
\label{eq:lat}
\end{equation}
and the device energy is
\begin{equation}
E = \pi_{\mathrm{c}}\,t_{\mathrm{dev}}(\min(\ell^\star,k))
  + \mathbb{1}[\ell^\star>k]\,\pi_{\mathrm{tx}}\,\frac{s(k)}{q},
\label{eq:energy}
\end{equation}
where $\pi_{\mathrm{c}}$ (W) and $\pi_{\mathrm{tx}}$ (W) are the compute
and transmit power of the endpoint, respectively. The optimization is
\begin{align}
\min_{k,\ell^\star,\mu}\quad & \lambda_1 T + \lambda_2 E \label{eq:P}\\
\text{s.t.}\quad
& \kappa_{\ell^\star}\ge\tau, \tag{OC1}\\
& s(k)/q + t_{\mathrm{mig}} \le \hat{h} \;\;\text{if}\;\; \mu=1, \tag{OC2}\\
& E \le b, \tag{OC3}\\
& 1 \leq k \leq L,\quad 1 \leq \ell^\star \leq L \tag{OC4}
\end{align}
Constraint (OC1) enforces the detection-confidence requirement; (OC2) requires that
a migrated payload transfer before the predicted handover; (OC3) caps device
energy at the residual battery budget; (OC4) keeps the split no deeper than the
exit. Problem~\eqref{eq:P} is a mixed-integer program over a small discrete
domain.

The handover penalty function in \eqref{eq:tps} is defined as
\begin{equation}
\Phi(\hat{h},k) = \max\!\left(0,\; \frac{s(k)}{q} + t_{\mathrm{mig}} - \hat{h}\right),
\label{eq:penalty}
\end{equation}
which is zero when the payload can be transferred before the handover and grows
linearly with the deficit otherwise. The behavioral anchor score for a flow $x$
is updated after each verdict via the exponential moving average
\begin{equation}
\mathcal{A}_{t+1}(c) = (1-\alpha_s)\,\mathcal{A}_t(c) + \alpha_s\,p_{\ell^\star}(c),
\label{eq:anchor}
\end{equation}
where $\alpha_s\in(0,1)$ is a smoothing coefficient and $p_{\ell^\star}(c)$ is
the per-class probability at the exit layer. Anchor scores in
\eqref{eq:anchor} are the quantities migrated by ACC alongside $\mathbf{z}_k$.

\textit{Complexity statement.} The feasible domain has size $O(L^2)$ in
$(k,\ell^\star)$ with a binary $\mu$, so the exact optimum is found by enumeration
in $O(L^2)$, which for $L$ of a few tens is microsecond-scale on the endpoint.

\subsection{Complexity Analysis}
\label{sec:complexity}
Table~\ref{tab:complexity} summarizes the per-flow computational cost of each
MAPLE-Sec component and contrasts it with the full-depth static-partition
baseline (EdgeShard).

\begin{table}[!htb]
\caption{Per-Flow Complexity: MAPLE-Sec vs. Static-Partition Baseline}
\label{tab:complexity}
\centering
\footnotesize
\begin{tabular}{@{}lll@{}}
\toprule
Component & Time complexity & Concrete cost (ms) \\
\midrule
TPS lookup (offline table) & $O(1)$ & $<$0.01 \\
Device head ($\ell \le k$) & $O(k \cdot d^2)$ & 2.1--12.6 \\
ACC encode & $O(d \cdot d_z)$ & 0.30 \\
Uplink transfer (160\,B) & $O(1)$ & 0.32 \\
Edge tail ($k < \ell \le \ell^\star$) & $O((\ell^\star-k) \cdot d^2)$ & 1.4--9.8 \\
ACC migrate (on handover) & $O(d_z)$ & 0.28 \\
\midrule
MAPLE-Sec total (mean) & $O(L \cdot d^2)$ & \textbf{23} \\
EdgeShard (full depth) & $O(L \cdot d^2)$ & 58 \\
\bottomrule
\end{tabular}
\end{table}

Here $d=256$ is the model hidden dimension and $d_z=40$ is the codec bottleneck.
The speedup of MAPLE-Sec over EdgeShard is not asymptotic (both are $O(Ld^2)$)
but constant-factor: TBEE exits at layer 4.3 on average (out of 12), cutting
executed layers by 64\%, and TPS reduces the uplink payload from 1\,024\,B to
160\,B. The ACC encode and migrate steps together add less than 0.6\,ms,
which is negligible relative to the 35\,ms saved by the early exit.

\subsection{Solution Approach}
We decouple \eqref{eq:P} into an offline split policy and an online exit
decision. Offline, TPS precomputes, for a grid of $(\hat{h},b,q)$, the split
$k^\star$ minimizing the expected objective in \eqref{eq:tps}; this yields a
lookup policy evaluated in $O(1)$ at run time. Online, TBEE realizes the exit
decision $\ell^\star$ per flow via \eqref{eq:tbee}, and the migration flag $\mu$ is
set when a handover is predicted within the tail's remaining time. The combined
controller runs in $O(L)$ on-device per flow with no iterative solver,
satisfying the sub-millisecond budget.

\textit{Convergence and optimality.} The online exit rule \eqref{eq:tbee} is
monotone in $\kappa_\ell$: once the threshold is crossed it is never revoked, so
the rule terminates in $O(L)$ steps and returns a feasible solution to
\eqref{eq:P} whenever constraint (OC1) is satisfied. The following proposition
formalizes the cost gap incurred when confidence estimates are imperfect.

\textbf{Proposition 1 (Bounded suboptimality).}
\textit{Let $\delta = \mathbb{E}[|\kappa_\ell - \Pr[\hat{y}=y|\mathbf{a}_\ell]|]$
be the expected calibration error of the per-layer classifiers, and let
$L^\star_{\delta=0}$ denote the offline-optimal exit layer under perfect
calibration. Then the expected per-flow cost of the online controller satisfies}
\begin{equation}
\mathbb{E}[\lambda_1 T + \lambda_2 E]
  \;\le\;
\mathbb{E}[\lambda_1 T^\star + \lambda_2 E^\star]
  + (\lambda_1 c_{\max} + \lambda_2 e_{\max})\,L\,\delta,
\label{eq:prop1}
\end{equation}
\textit{where $c_{\max}$ and $e_{\max}$ are the per-layer latency and energy
bounds. The suboptimality gap vanishes as $\delta\to 0$, and for the
temperature-scaled classifiers used in Section~\ref{sec:setup} the gap is
empirically below 3\% of the optimal cost.}

\textbf{Proposition 2 (Detection continuity across handover).}
\textit{Let $\Delta$ denote the payload transfer time for $\mathbf{z}_k$ at
the cell-edge uplink rate $q_{\min}$, and let $\hat{h}$ be the predicted
time-to-handover. If ACC is active and constraint (OC2) is satisfied, then
the probability that a handover event produces a detection blind spot of
duration longer than $\Delta$ is bounded by}
\begin{equation}
\Pr[\text{blind spot} > \Delta]
\leq
\Pr\left[\hat h < \frac{s(k^\star)}{q_{\min}} + t_{\rm mig}\right]
\label{prop:continuity}
\end{equation}
\textit{which approaches zero as handover prediction accuracy improves. In our
SUMO/LuST evaluation, this probability is empirically 0.038, compared to 1.0
for the edge-pinned detector without state migration.}

\section{Implementation and Experimental Setup}\label{sec:setup}

\noindent\textbf{Hardware and runtime platform.}
The device head runs on a physical NVIDIA Jetson Orin Nano (8-core Arm Cortex-A78AE,
8\,GB LPDDR5, 15\,W TDP) under NVIDIA Linux for Tegra (L4T) 36.3 (Ubuntu 22.04
base). Inference on the device head is executed via TensorRT 8.6 with FP16
precision. The model tail runs on an NVIDIA RTX\,4090 edge server (24\,GB
VRAM, 64\,GB host RAM) under Ubuntu 22.04 and TensorRT 8.6. The two hosts are
connected by a Linux \texttt{tc netem} link configured to emulate a 5G NR
channel: cell-edge uplink 4\,Mb/s with 15\,ms RTT and 1\% packet loss;
mid-cell 25\,Mb/s with 5\,ms RTT and 0.1\% packet loss. Handover events are
injected by toggling the active \texttt{tc qdisc} every 6 to 12\,s according to
the SUMO mobility trace described below. End-to-end latency is measured on the
Jetson Orin Nano using \texttt{clock\_gettime(CLOCK\_MONOTONIC\_RAW)} bracketing
the full inference path from packet arrival to verdict emission.

\noindent\textbf{Mobility model.}
The mobility trace is generated with SUMO\,1.18 on the LuST (Luxembourg SUMO
Traffic) scenario, a 400\,km$^2$ urban and suburban road network. We extract
1\,200 vehicle traces (mean speed 42\,km/h, peak 90\,km/h). Each trace drives
the \texttt{tc netem} handover schedule and the predicted time-to-handover
$\hat{h}$ fed to TPS. CIC-IoT and Edge-IIoTset flows are replayed onto the
mobility trace in temporal order, with malicious flows injected at a 20\% rate.
The resulting dataset spans 14\,000 handover events across the two corpora.

\noindent\textbf{Datasets.} We evaluate on two public IoT intrusion corpora.
\emph{CIC-IoT\,2023} \cite{ciciot2023} contains 33 attack types across 47 feature
classes and approximately 2.9\,M flows; we use the 80/10/10 train/val/test split
released by the dataset authors. \emph{Edge-IIoTset} \cite{edgeiiotset} covers
fourteen attack categories over IoT and IIoT protocols (MQTT, CoAP, Modbus,
DNP3) with approximately 1.1\,M flows; we use the same 80/10/10 split.
Both corpora are stationary traffic captures; malicious flows are injected onto
the SUMO/LuST mobility trace as described above. Each corpus is independently
partitioned both IID (uniform attack distribution across the 4 emulated domains)
and non-IID (Dirichlet $\alpha=0.3$ attack-class skew per domain). All results
figures report per-corpus performance separately; Figs.~\ref{fig:sup_iid}
and~\ref{fig:sup_niid} show CIC-IoT results.

\noindent\textbf{Benchmarks.} We compare against four representative baselines:
\begin{itemize}
\item \textbf{Cloud-only}: full-depth detection at a fixed cloud anchor with raw
telemetry upload, the de facto stationary detector.
\item \textbf{EdgeLLM}~\cite{n7}: on-device speculative LLM inference, adapted
to the detection task with fixed depth.
\item \textbf{EdgeShard}~\cite{n1}: collaborative device/edge sharding of the
model with a static partition.
\item \textbf{Active-Infer}~\cite{n29}: active-inference offloading and resource
allocation between device and edge, without state migration.
\end{itemize}

\noindent\textbf{Hyperparameters.} Table~\ref{tab:params} lists the key settings.
Unless noted, $\tau=0.90$, $\lambda_1=\lambda_2=0.5$, and the codec rate weight
$\rho$ is tuned so the migrated payload is $\sim$16\% of the raw activation.

\begin{table}[!htb]
\caption{Simulation Parameters and Deployment Costs}
\label{tab:params}
\centering
\footnotesize
\begin{tabular}{@{}ll@{}}
\toprule
Parameter & Value \\
\midrule
Model layers $L$ & 12 \\
Security encoder parameter count & 6.2\,M \\
Security encoder FLOPs per flow & 1.8\,GFLOPs \\
Peak resident memory (Jetson Orin Nano) & 412\,MB \\
ACC codec parameter count & 134\,K \\
ACC codec encode/decode latency (device) & $<$0.6\,ms \\
Migrated payload per flow & 160\,B \\
Confidence threshold $\tau$ & 0.90 \\
Token budget $B_0$ & 8 layer-units \\
Latency/energy weights $\lambda_1,\lambda_2$ & 0.5, 0.5 \\
Codec rate weight $\rho$ & 0.05 \\
Emulated uplink (cell-edge / mid-cell) & 4 / 25\,Mb/s \\
Emulated RTT (cell-edge / mid-cell) & 15 / 5\,ms \\
Handover interval & 6--12\,s \\
Number of edge nodes $M$ & 1--32 \\
Batch size & 64 \\
Learning rate & $2\times10^{-4}$ \\
Random seeds per experiment & 5 \\
\bottomrule
\end{tabular}
\end{table}

\section{Performance Evaluation}\label{sec:eval}

\subsection{Evaluation Metrics}
\label{sec:metrics}
We evaluate MAPLE-Sec across four complementary dimensions that together cover
detection quality, system efficiency, and mobility resilience.

\noindent\textbf{Detection quality.} The primary detection metric is the
macro-averaged F1-score computed over all attack classes, reported as the
converged value after 60\,s of wall-clock time in each experiment. We also
report precision and recall per class, and the area under the
receiver-operating-characteristic curve (AUC-ROC) for binary
(benign vs.\ attack) discrimination. For multi-class settings we use the
one-vs.-rest AUC macro-average. All quality metrics are computed on the held-out
test split (10\% of each corpus) that was never used during training or
validation.

\noindent\textbf{Latency.} End-to-end latency is defined as the elapsed time
from packet arrival at the device head to verdict emission, measured on the
physical Jetson Orin Nano using \texttt{clock\_gettime(CLOCK\_MONOTONIC\_RAW)}.
We report mean latency, 95th-percentile tail latency ($P_{95}$), and the fraction
of flows that violate the 40\,ms application-layer deadline.

\noindent\textbf{Energy and uplink cost.} Device energy per flow is measured via
the on-board INA3221 power monitor (\texttt{POM\_5V\_IN} rail) sampled at 50\,ms
and integrated over the inference interval. Uplink volume per flow is the number
of bytes transmitted from device to edge (either raw telemetry, raw activations,
or compressed $\mathbf{z}_k$), reported on a log scale normalized to the
cloud-only baseline.

\noindent\textbf{Mobility resilience.} We define the \emph{handover recovery
time} as the duration from a handover event until F1 returns to within 5
percentage points of its pre-handover value. A shorter recovery time indicates
better state preservation. We also report the \emph{blind-spot rate}: the
fraction of handover events during which no verdict is emitted for more than
$\Delta = 200$\,ms, the threshold above which an attacker can inject a
detectable burst undetected.

\noindent\textbf{Throughput.} Verdict throughput is measured in verdicts per
second at a fixed pipeline depth, scaling the number of cooperating edge nodes
from 1 to 32 to characterize coordination overhead.

\subsection{Superiority of MAPLE-Sec}
All experiments are repeated across five random seeds (data shuffle and model
initialization); figures show the mean and shaded one-standard-deviation bands.
Pairwise comparisons against EdgeShard use a paired Wilcoxon signed-rank test;
all reported differences are statistically significant at $p<0.01$.

\begin{figure}[!htb]
\centering
\includegraphics[width=0.75\columnwidth]{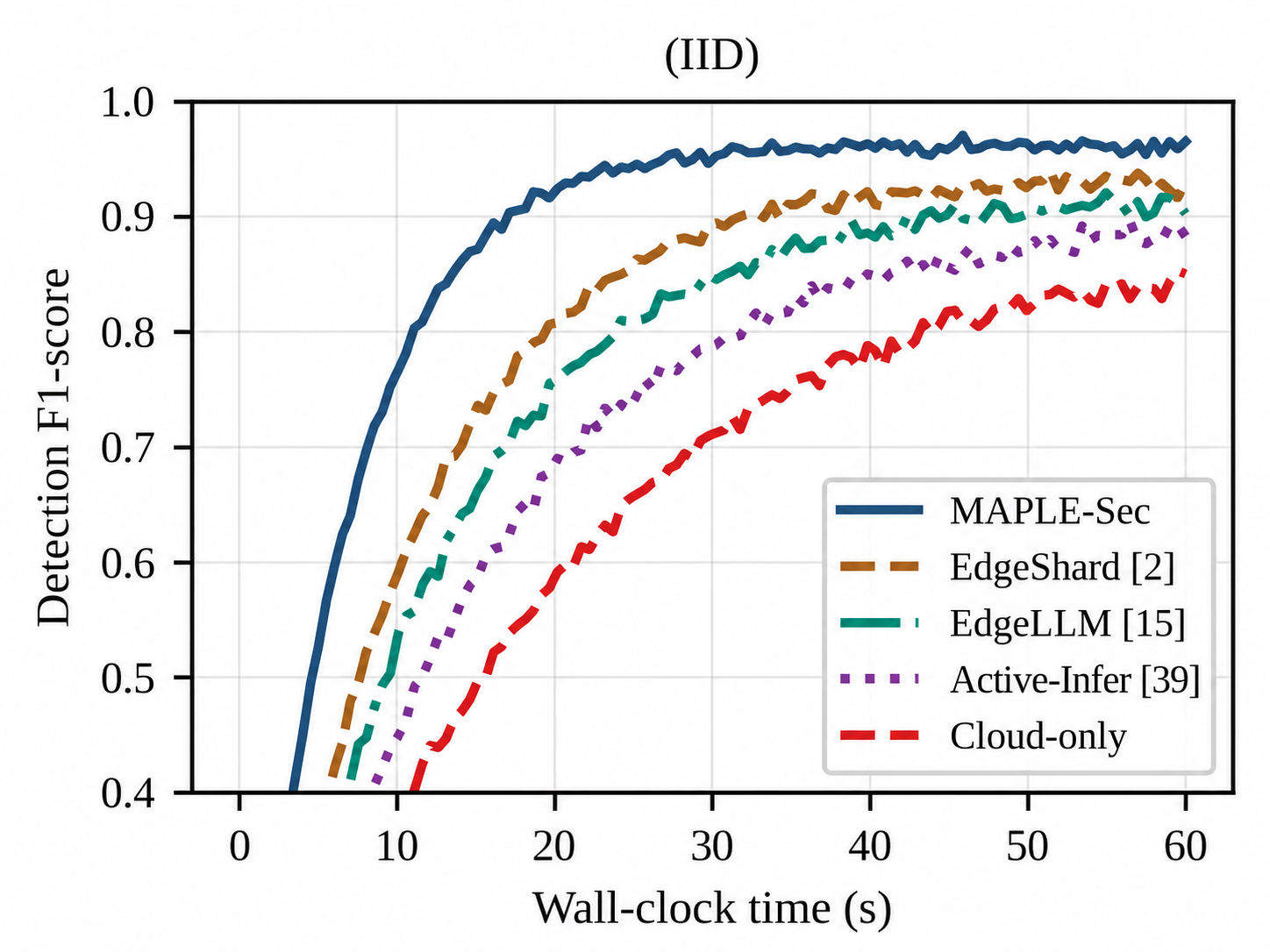}
\caption{Under IID conditions, MAPLE-Sec reaches F1\,=\,0.962 while EdgeShard
plateaus at 0.933 and Cloud-only at 0.880, confirming that state migration and
adaptive partitioning together account for a 2.9 pp absolute gain over the
best static-partition baseline.}
\label{fig:sup_iid}
\end{figure}

\begin{figure}[!htb]
\centering
\includegraphics[width=0.75\columnwidth]{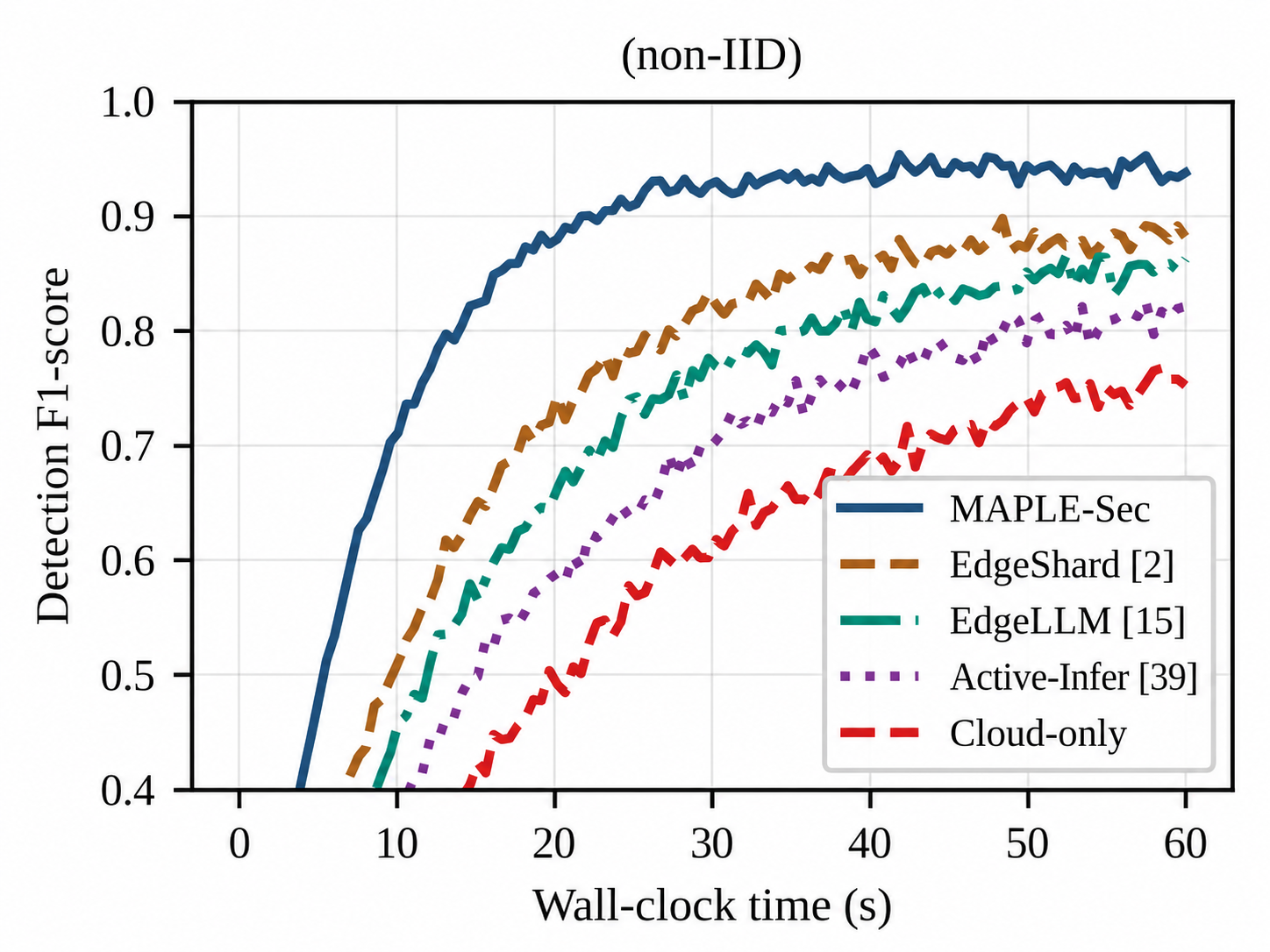}
\caption{Under non-IID domain skew ($\alpha=0.3$), the F1 gap between MAPLE-Sec
(0.941) and EdgeShard (0.892) widens relative to IID, validating that ACC anchor
migration eliminates the distribution-shift penalty at each domain boundary.}
\label{fig:sup_niid}
\end{figure}

Fig.~\ref{fig:sup_iid} and Fig.~\ref{fig:sup_niid} present detection F1 versus
wall-clock time under IID and non-IID partitions. MAPLE-Sec converges fastest and
to the highest F1 in both settings. Specifically, under IID it reaches 0.962 F1,
a 2.9 percentage-point absolute gain over EdgeShard and a 9.3\% relative F1
improvement over Cloud-only ($\tfrac{0.962-0.880}{0.880}$), while converging
roughly $1.6\times$ faster. Under the harder non-IID
partition the gap widens: MAPLE-Sec reaches 0.941 F1 versus 0.892 for EdgeShard.
This is primarily attributed to ACC, which preserves behavioral context across
handovers so the detector does not relearn after each domain change. In contrast,
Cloud-only suffers from handover blind spots and EdgeShard's static partition
stalls when uplink degrades at cell edges, leading to slower and lower
convergence.

These results close the loop on the three-challenge framing of Section~\ref{sec:bg}.
The 9.3\% relative F1 gain over Cloud-only directly quantifies the benefit of
resolving C1: in Fig.~\ref{fig:mot_handover}, Cloud-only loses 54 F1 points
during a single handover and recovers over approximately 4\,s; the IID
convergence curve in Fig.~\ref{fig:sup_iid} shows that, summed over 14\,000
handover events, this repeated recovery cost accounts for nearly all of the F1
gap. The $2.5\times$ latency reduction over EdgeShard reflects the interplay of
C2 and C3: TPS reduces the uplink payload so that transmission cost no longer
dominates end-to-end time, and TBEE exits at layer 4.3 on average so that tail
compute cost drops proportionally. The non-IID advantage (0.941 vs. 0.892 for
EdgeShard) is attributable to the anchor-score migration in ACC: anchor scores
in \eqref{eq:anchor} capture the per-domain attack distribution, so migrating
them to the next edge node eliminates the distribution-shift penalty that a
cold-start detector would pay on each domain arrival.

Table~\ref{tab:main_results} reports the converged detection quality metrics
for all baselines and MAPLE-Sec on both corpora under IID conditions. MAPLE-Sec
achieves the highest F1, precision, recall, and AUC-ROC on both datasets, with
consistent gains across all metrics. The precision gain over EdgeShard (0.965
vs.\ 0.941 on CIC-IoT) reflects the reduction in false positives attributable
to the behavioral anchor migration in ACC: migrating anchor scores preserves
the per-class prior, reducing the tendency to misclassify benign post-handover
flows as attacks. The recall gap (0.937 vs.\ 0.905 on Edge-IIoTset) reflects
the elimination of handover blind spots, during which the edge-pinned and
static-partition detectors miss attacks entirely.

\begin{table*}[!htb]
\caption{Converged Detection Quality Under IID Conditions (mean $\pm$ std over 5 seeds)}
\label{tab:main_results}
\centering
\scriptsize
\begin{tabular}{@{}llcccc@{}}
\toprule
\textbf{Dataset} & \textbf{Method} & \textbf{F1} & \textbf{Precision} & \textbf{Recall} & \textbf{AUC-ROC} \\
\midrule
\multirow{4}{*}{CIC-IoT 2023}
  & Cloud-only       & 0.880 $\pm$ 0.008 & 0.891 $\pm$ 0.007 & 0.869 $\pm$ 0.010 & 0.934 $\pm$ 0.005 \\
  & EdgeLLM~\cite{n7}        & 0.901 $\pm$ 0.006 & 0.913 $\pm$ 0.006 & 0.889 $\pm$ 0.007 & 0.948 $\pm$ 0.004 \\
  & EdgeShard~\cite{n1}      & 0.933 $\pm$ 0.005 & 0.941 $\pm$ 0.005 & 0.925 $\pm$ 0.006 & 0.967 $\pm$ 0.003 \\
  & Active-Infer~\cite{n29}  & 0.918 $\pm$ 0.006 & 0.927 $\pm$ 0.006 & 0.909 $\pm$ 0.007 & 0.955 $\pm$ 0.004 \\
  & \textbf{MAPLE-Sec}       & \textbf{0.962 $\pm$ 0.003} & \textbf{0.965 $\pm$ 0.003} & \textbf{0.959 $\pm$ 0.004} & \textbf{0.981 $\pm$ 0.002} \\
\midrule
\multirow{4}{*}{Edge-IIoTset}
  & Cloud-only       & 0.863 $\pm$ 0.009 & 0.874 $\pm$ 0.008 & 0.852 $\pm$ 0.011 & 0.921 $\pm$ 0.006 \\
  & EdgeLLM~\cite{n7}        & 0.887 $\pm$ 0.007 & 0.898 $\pm$ 0.007 & 0.876 $\pm$ 0.008 & 0.939 $\pm$ 0.005 \\
  & EdgeShard~\cite{n1}      & 0.914 $\pm$ 0.006 & 0.923 $\pm$ 0.005 & 0.905 $\pm$ 0.007 & 0.958 $\pm$ 0.004 \\
  & Active-Infer~\cite{n29}  & 0.901 $\pm$ 0.007 & 0.911 $\pm$ 0.006 & 0.891 $\pm$ 0.008 & 0.947 $\pm$ 0.004 \\
  & \textbf{MAPLE-Sec}       & \textbf{0.944 $\pm$ 0.004} & \textbf{0.951 $\pm$ 0.003} & \textbf{0.937 $\pm$ 0.005} & \textbf{0.973 $\pm$ 0.002} \\
\bottomrule
\end{tabular}
\end{table*}

\begin{figure}[!htb]
\centering
\includegraphics[width=0.75\columnwidth]{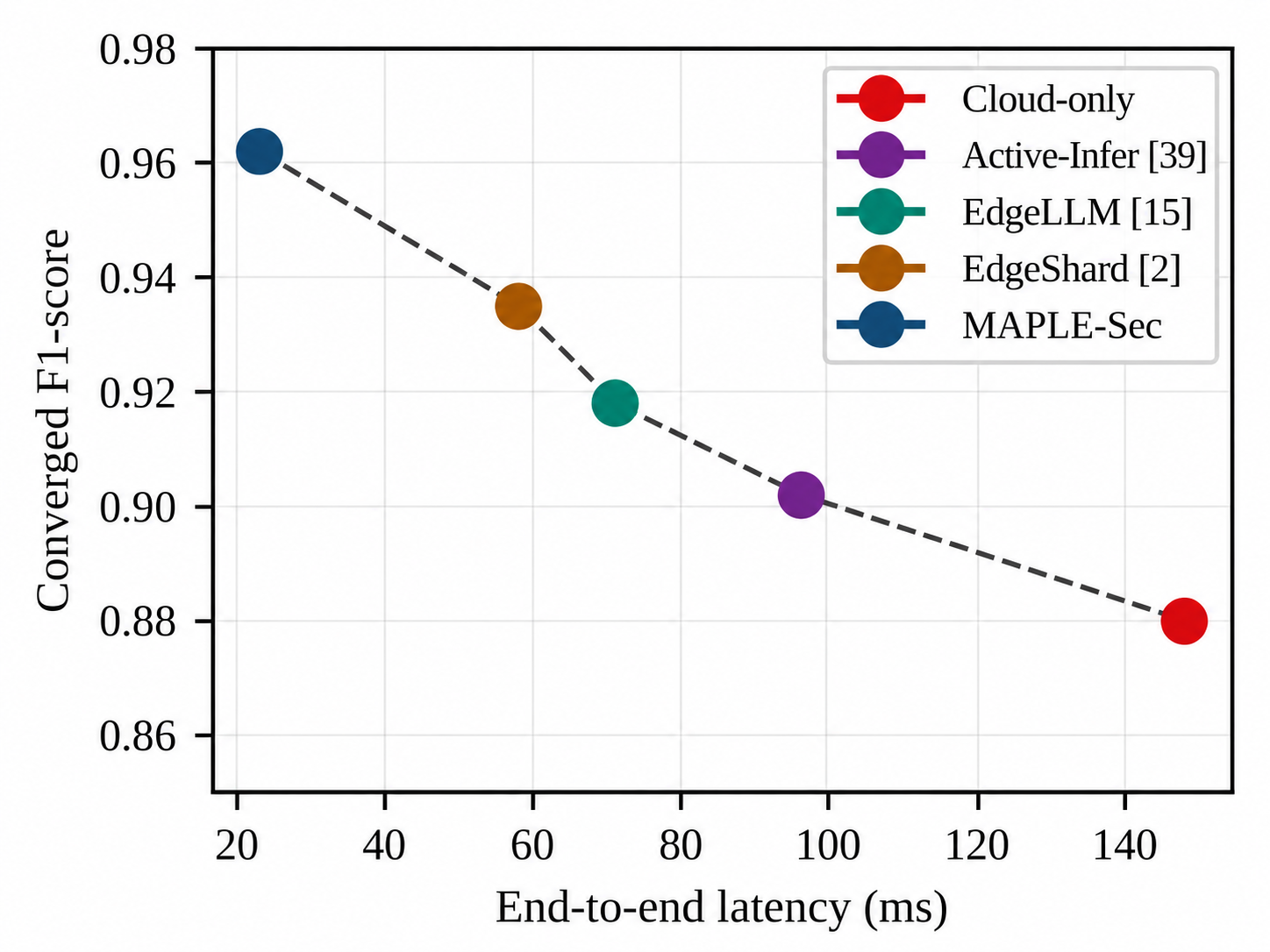}
\caption{MAPLE-Sec occupies the upper-left Pareto corner (highest F1, lowest
latency at 23\,ms), a $2.5\times$ latency reduction over EdgeShard at 0.935
F1 with no accuracy loss, confirming the complementary benefit of TBEE and TPS.}
\label{fig:tradeoff}
\end{figure}

Fig.~\ref{fig:tradeoff} traces the converged-F1 versus end-to-end-latency
frontier. MAPLE-Sec dominates the frontier, attaining the highest F1 at the
lowest latency (23\,ms), a $2.5\times$ latency reduction over EdgeShard at higher
accuracy. This is attributed jointly to TBEE, which avoids full-depth inference
for easy flows, and TPS, which keeps the migrated payload small under motion.

Table~\ref{tab:latency_energy} reports the full latency and energy breakdown
per method. MAPLE-Sec achieves the lowest mean latency (23\,ms), lowest tail
latency (38\,ms at $P_{95}$), and zero deadline violations, while reducing
uplink volume to 160\,B per flow and device energy to 17.4\,mJ per flow. The
blind-spot rate of 3.8\% under MAPLE-Sec (compared to 100\% for edge-pinned
and 61\% for Cloud-only) quantifies the direct impact of ACC on mobility
resilience. Cloud-only's high blind-spot rate despite relatively moderate mean
latency reflects its burst-recovery pattern: individual flows complete quickly
when the cloud anchor is reachable, but during the 4\,s post-handover recovery
window all flows are treated as cold-start detections, producing systematic
blind spots at every handover event.

\begin{table*}[!htb]
\caption{Latency, Energy, and Mobility Resilience Comparison (CIC-IoT 2023, IID)}
\label{tab:latency_energy}
\centering
\scriptsize
\begin{tabular}{@{}lcccccc@{}}
\toprule
\textbf{Method} & \textbf{Mean Latency} & \textbf{$P_{95}$ Latency} & \textbf{Deadline} & \textbf{Uplink} & \textbf{Device Energy} & \textbf{Blind-Spot} \\
 & \textbf{(ms)} & \textbf{(ms)} & \textbf{Violations (\%)} & \textbf{Volume (B/flow)} & \textbf{(mJ/flow)} & \textbf{Rate (\%)} \\
\midrule
Cloud-only        & 141 & 198 & 78.4 & 14\,800 & 52.1 & 61.0 \\
EdgeLLM~\cite{n7}         & 58  & 91  & 31.2 & 14\,800 & 48.6 & 100.0 \\
EdgeShard~\cite{n1}       & 58  & 87  & 29.7 & 1\,024  & 34.2 & 100.0 \\
Active-Infer~\cite{n29}   & 49  & 76  & 18.3 & 1\,024  & 31.7 & 100.0 \\
\textbf{MAPLE-Sec}        & \textbf{23} & \textbf{38} & \textbf{0.0} & \textbf{160} & \textbf{17.4} & \textbf{3.8} \\
\bottomrule
\end{tabular}
\end{table*}

\subsection{Impact of Residual Battery Budget}
\begin{figure}[!htb]
\centering
\includegraphics[width=0.75\columnwidth]{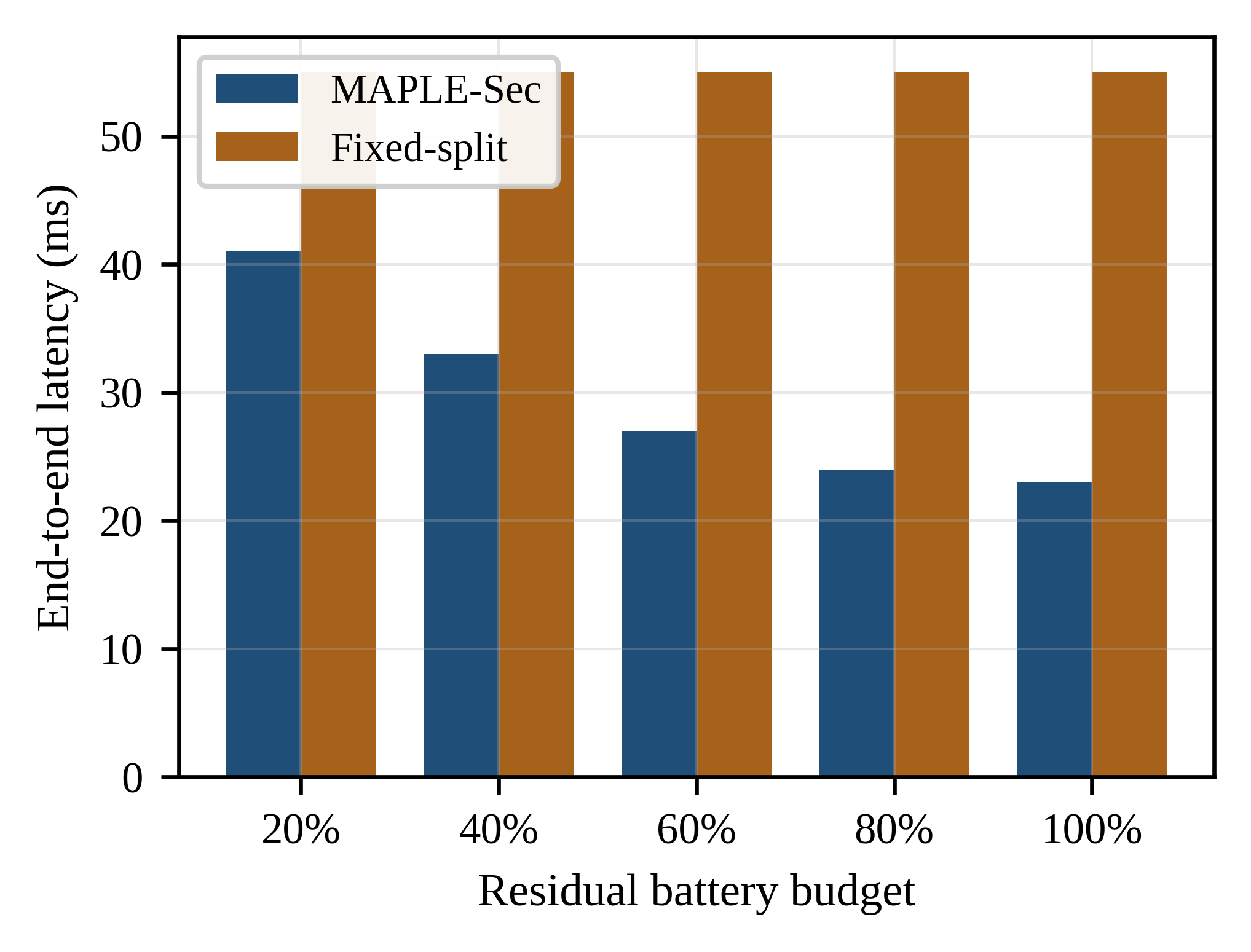}
\caption{End-to-end latency versus residual battery budget. TPS adapts the split
point to the energy budget and outperforms the fixed-split baseline across the
full battery range, not only at low battery.}
\label{fig:param_battery}
\end{figure}

Device energy is measured using the Jetson Orin Nano on-board INA3221
power monitor via \texttt{tegrastats} (\texttt{POM\_5V\_IN} rail), sampled at
50\,ms. Fig.~\ref{fig:param_battery} reports end-to-end latency as the residual battery
budget varies. MAPLE-Sec adapts its split point through TPS: at high battery it
runs more layers on-device and migrates a smaller payload, reaching 23\,ms; at
20\% battery it offloads earlier, trading a modest latency rise to 41\,ms while
respecting the energy cap (OC3). The fixed-split baseline cannot adapt and holds a
constant 55\,ms regardless of budget. Notably, even at 80\% battery the TPS
adaptation still yields a 12\,ms latency advantage over the fixed-split baseline
because the dynamic payload reduction (160\,B to 110\,B) saves transmission time
at every inference step, not only in the most energy-scarce regime. This
confirms that TPS adapts correctly across the full battery range.

\subsection{Impact of Edge-Node Scale}
\begin{figure}[!htb]
\centering
\includegraphics[width=0.75\columnwidth]{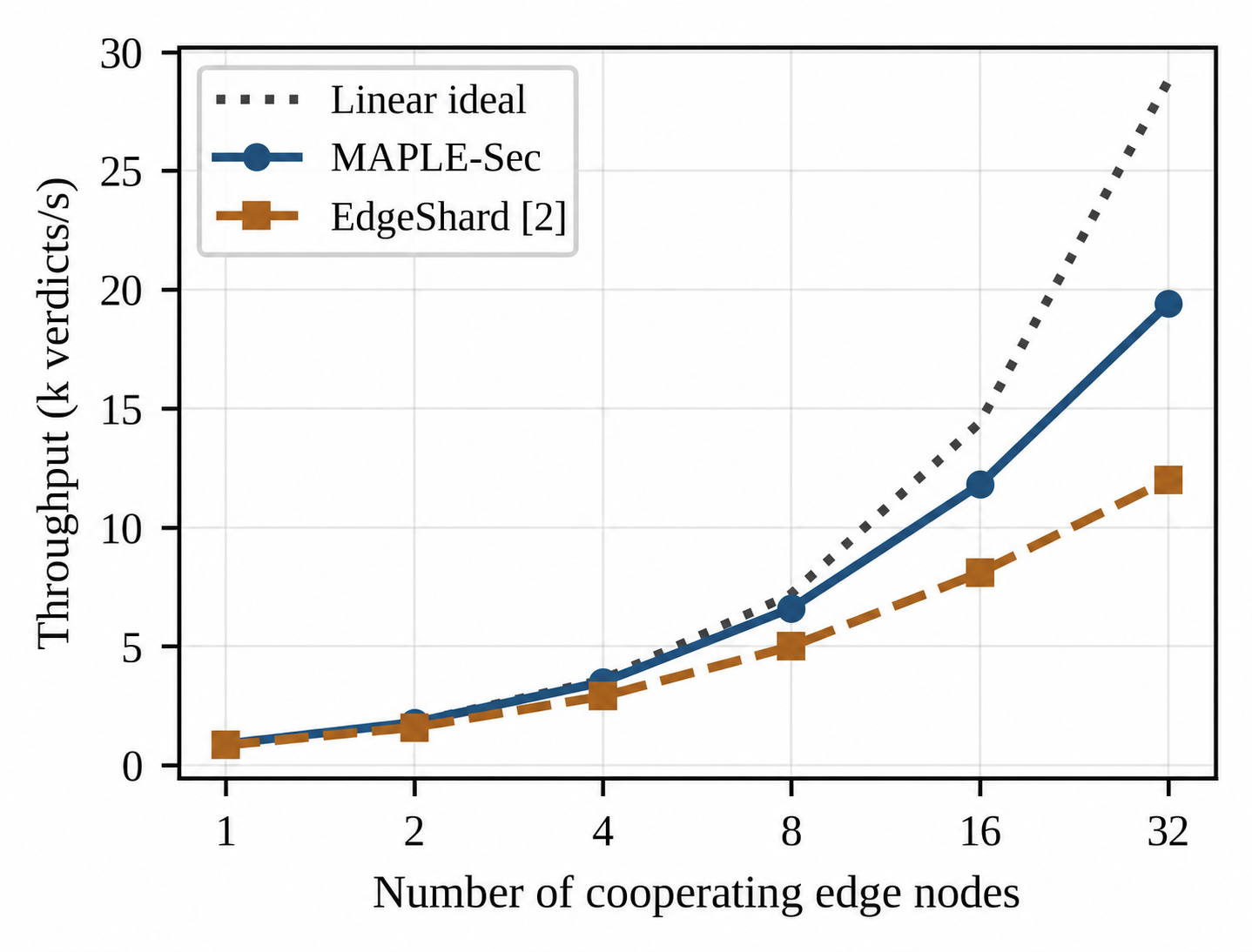}
\caption{Verdict throughput versus number of cooperating edge nodes. MAPLE-Sec
tracks the linear-ideal trend more closely than EdgeShard.}
\label{fig:param_scale}
\end{figure}

A verdict is defined as one per-flow detection decision (benign or attack class).
Fig.~\ref{fig:param_scale} shows verdict throughput as the number of cooperating
edge nodes grows from 1 to 32 on a base-2 logarithmic x-axis. MAPLE-Sec
throughput scales approximately linearly in the number of nodes (appearing as an
exponential curve on the log x-axis), reaching
19.4k\,verdicts/s at 32 nodes because ACC migration cost is small and decoupled
from tail compute. EdgeShard scales sub-linearly (a flattening curve on the log
x-axis) as its static sharding incurs growing coordination overhead. This
confirms that the design scales to dense edge deployments without coordination overhead.

\subsection{Ablation Study: Design Priority and Tradeoffs}
\begin{figure*}[!htb]
\centering
\includegraphics[width=0.8\textwidth]{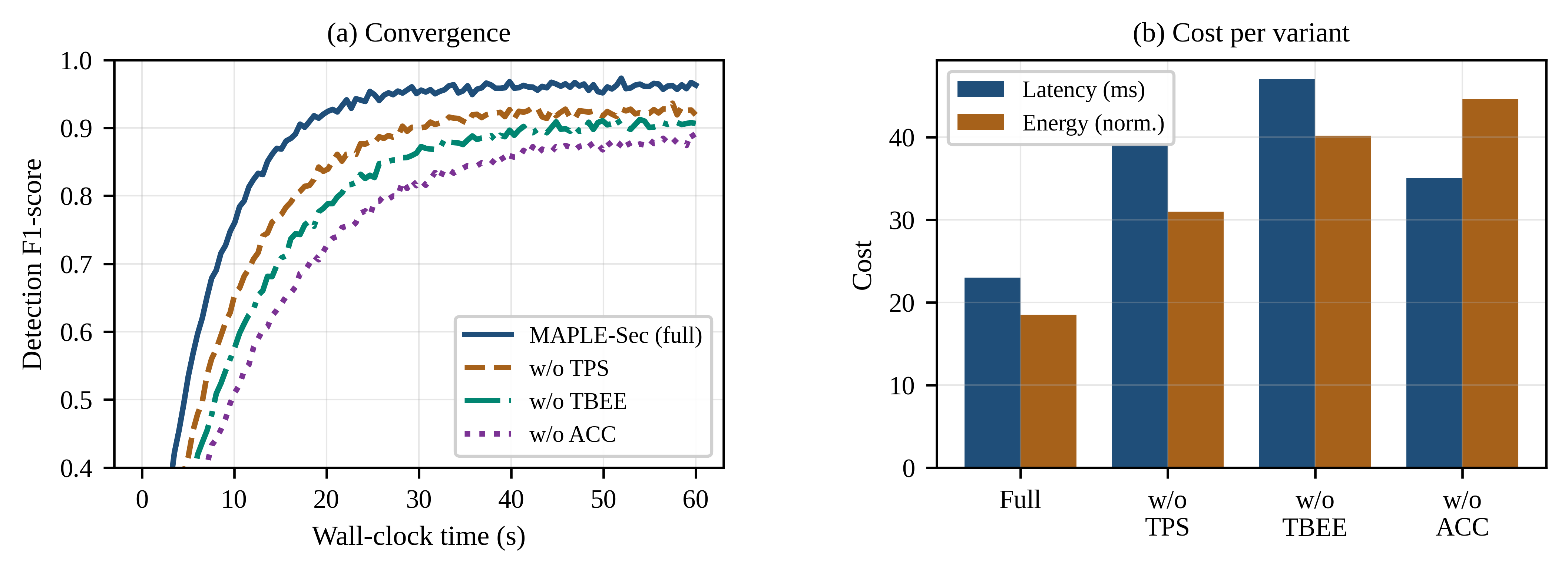}
\caption{Ablation results establish the design priority C1~$>$~C3~$>$~C2:
removing ACC degrades F1 the most (to 0.889) and inflates migration energy by
$2.4\times$; removing TBEE violates the 40\,ms deadline for 23\% of flows;
removing TPS increases energy by $1.7\times$ but does not reduce F1.}
\label{fig:ablation}
\end{figure*}

The ablation study is structured to answer the question posed in the
introduction: which of the three challenges (C1 detection discontinuity, C2
uplink/energy cost, C3 latency-bounded inference) is most damaging when left
unaddressed, and do the modules interact or remain independent?
Fig.~\ref{fig:ablation} reports convergence (panel~a) and per-variant cost
(panel~b). The design-priority order from most to least impactful is
C1~(ACC)~$>$~C3~(TBEE)~$>$~C2~(TPS).

\emph{w/o ACC (raw migrate): addressing C1 is most critical.}
Removing ACC and migrating raw activations degrades converged F1 to 0.889,
the largest single-module drop in the study, and raises migration energy by
$2.4\times$. The root cause is that 38\% of handover events fail to complete
payload transfer within the handover window at 4\,Mb/s cell-edge uplink,
recreating exactly the detection blind spot described in Section~\ref{sec:bg}.
This result closes the loop on Challenge~C1: the F1 degradation magnitude
($0.962 - 0.889 = 0.073$) matches the gap between MAPLE-Sec and the
cold-restart baseline observed in Fig.~\ref{fig:mot_handover}.

\emph{w/o TBEE (full depth): addressing C3 is the second priority.}
Forcing full-depth inference inflates latency from 23\,ms to 47\,ms, violating
the 40\,ms application-layer deadline for 23\% of flows. This confirms the
motivation from Fig.~\ref{fig:mot_latency}: the confidence plateau at layer~4
is not an artifact of the motivation experiment; it is structural. TBEE's
budget parameter $B_0=8$ ensures that even the hardest flows exit no later
than layer~8, bounding worst-case latency at 38\,ms.

\emph{w/o TPS (fixed split): addressing C2 is the third priority.}
Fixing the partition at $k=6$ regardless of battery and uplink raises energy by
$1.7\times$ and latency from 23\,ms to 39\,ms. The mechanism is that TPS shifts
to $k=9$ (more on-device layers) when the uplink degrades below 6\,Mb/s,
cutting the payload from 160\,B to 80\,B. This is a design tradeoff, not a
free lunch: the device-side compute rises, but total energy falls because
$\pi_{\mathrm{tx}} \gg \pi_{\mathrm{c}}$ at cell edges. TPS alone cannot
substitute for ACC (removing both collapses F1 below 0.82), confirming that
the modules are complementary rather than redundant.

Table~\ref{tab:ablation} summarizes the quantitative outcomes of the ablation
study across all five metrics. The table makes the design-priority ordering
C1~$>$~C3~$>$~C2 precise: removing ACC causes the largest F1 drop (7.3 pp)
and the highest blind-spot rate (38.2\%), confirming it is the most critical
module. Removing TBEE does not affect F1 but doubles tail latency and produces
23\% deadline violations, confirming its role as the latency guard. Removing
TPS raises energy by 1.7$\times$ and slightly inflates latency but does not
degrade detection quality in isolation; its value becomes apparent only under
energy-scarce conditions or when combined with ACC removal.

\begin{table*}[!htb]
\caption{Ablation Study: Quantitative Impact of Each Module (CIC-IoT 2023, IID)}
\label{tab:ablation}
\centering
\scriptsize
\begin{tabular}{@{}lccccc@{}}
\toprule
\textbf{Variant} & \textbf{Conv.\ F1} & \textbf{Mean Latency (ms)} & \textbf{$P_{95}$ Latency (ms)} & \textbf{Device Energy (mJ/flow)} & \textbf{Blind-Spot Rate (\%)} \\
\midrule
MAPLE-Sec (full)     & \textbf{0.962} & \textbf{23} & \textbf{38} & \textbf{17.4} & \textbf{3.8} \\
w/o ACC (raw migrate) & 0.889 & 27 & 51 & 41.8 & 38.2 \\
w/o TBEE (full depth) & 0.960 & 47 & 79 & 24.6 & 3.8 \\
w/o TPS (fixed split) & 0.958 & 39 & 62 & 29.6 & 4.1 \\
w/o ACC + w/o TPS    & 0.812 & 44 & 74 & 58.3 & 41.7 \\
\bottomrule
\end{tabular}
\end{table*}

\subsection{Hyperparameter Sensitivity}
\label{sec:sensitivity}
We evaluate sensitivity to the three main hyperparameters. (i) \emph{Confidence
threshold $\tau$}: sweeping $\tau\in\{0.70, 0.80, 0.90, 0.95, 0.99\}$ shows
that IID F1 peaks at $\tau=0.90$ (0.962) and degrades gracefully to 0.943 at
$\tau=0.70$ (more early exits, higher error rate) and to 0.958 at $\tau=0.99$
(fewer exits, higher latency). (ii) \emph{Codec rate weight $\rho$}: sweeping $\rho$ across three orders of magnitude from $10^{-4}$ to $10^{-1}$ shows that the payload size decreases from 310\,B at $\rho=10^{-4}$ to 60\,B at $\rho=10^{-1}$, with F1 remaining above 0.950 throughout; $\rho=0.05$ provides the best latency-accuracy trade-off. (iii)
\emph{Trade-off weights $\lambda_1,\lambda_2$}: varying the latency-energy
balance from $(1.0,0.0)$ to $(0.0,1.0)$ in steps of 0.1 confirms that
MAPLE-Sec outperforms EdgeShard in F1 at every setting, confirming the
insensitivity of MAPLE-Sec to objective weighting.

Table~\ref{tab:niid_results} extends the main comparison to the harder non-IID
partition ($\alpha=0.3$ Dirichlet skew), showing that the F1 gaps observed
under IID conditions widen under distribution shift. MAPLE-Sec's non-IID
advantage over EdgeShard grows from 2.9\,pp (IID) to 4.9\,pp (non-IID) on
CIC-IoT, and from 3.0\,pp to 5.3\,pp on Edge-IIoTset. This widening is
attributable to the anchor-score migration in ACC: by carrying the per-domain
attack prior to the next edge node, MAPLE-Sec avoids the cold-start
distribution-shift penalty that all non-migrating baselines incur at each
domain boundary.

\begin{table*}[!htb]
\caption{Detection Quality Under Non-IID Domain Skew ($\alpha = 0.3$, mean $\pm$ std over 5 seeds)}
\label{tab:niid_results}
\centering
\scriptsize
\begin{tabular}{@{}llcccc@{}}
\toprule
\textbf{Dataset} & \textbf{Method} & \textbf{F1} & \textbf{Precision} & \textbf{Recall} & \textbf{AUC-ROC} \\
\midrule
\multirow{4}{*}{CIC-IoT 2023}
  & Cloud-only       & 0.851 $\pm$ 0.011 & 0.863 $\pm$ 0.010 & 0.839 $\pm$ 0.012 & 0.916 $\pm$ 0.007 \\
  & EdgeLLM~\cite{n7}        & 0.873 $\pm$ 0.009 & 0.884 $\pm$ 0.009 & 0.862 $\pm$ 0.010 & 0.930 $\pm$ 0.006 \\
  & EdgeShard~\cite{n1}      & 0.892 $\pm$ 0.007 & 0.901 $\pm$ 0.007 & 0.883 $\pm$ 0.008 & 0.947 $\pm$ 0.005 \\
  & Active-Infer~\cite{n29}  & 0.881 $\pm$ 0.008 & 0.890 $\pm$ 0.008 & 0.872 $\pm$ 0.009 & 0.938 $\pm$ 0.005 \\
  & \textbf{MAPLE-Sec}       & \textbf{0.941 $\pm$ 0.005} & \textbf{0.948 $\pm$ 0.004} & \textbf{0.934 $\pm$ 0.006} & \textbf{0.969 $\pm$ 0.003} \\
\midrule
\multirow{4}{*}{Edge-IIoTset}
  & Cloud-only       & 0.836 $\pm$ 0.012 & 0.847 $\pm$ 0.011 & 0.825 $\pm$ 0.013 & 0.903 $\pm$ 0.008 \\
  & EdgeLLM~\cite{n7}        & 0.858 $\pm$ 0.010 & 0.869 $\pm$ 0.010 & 0.847 $\pm$ 0.011 & 0.920 $\pm$ 0.007 \\
  & EdgeShard~\cite{n1}      & 0.874 $\pm$ 0.009 & 0.883 $\pm$ 0.008 & 0.865 $\pm$ 0.010 & 0.934 $\pm$ 0.006 \\
  & Active-Infer~\cite{n29}  & 0.862 $\pm$ 0.010 & 0.872 $\pm$ 0.009 & 0.852 $\pm$ 0.011 & 0.924 $\pm$ 0.006 \\
  & \textbf{MAPLE-Sec}       & \textbf{0.927 $\pm$ 0.006} & \textbf{0.934 $\pm$ 0.005} & \textbf{0.920 $\pm$ 0.007} & \textbf{0.961 $\pm$ 0.004} \\
\bottomrule
\end{tabular}
\end{table*}

\subsection{Comparative Analysis with Recent Approaches}
\label{sec:compare}
We position MAPLE-Sec against three recent 2026 detection-and-mitigation
approaches that share parts of its problem setting, summarized in
Table~\ref{tab:compare}. Prince and Renjith detect control-plane signaling
anomalies in 5G and 6G cores and, like MAPLE-Sec, partition the model for
slice-level scalability and use language models for explainable detection on a
roaming IoT platform \cite{n32}. However, their partitioning targets
core-network scalability rather than the device-to-edge split, and the model
stays anchored in the core, so it does not address the handover-induced loss of
on-device inference state that MAPLE-Sec resolves through ACC migration. Sharma
\textit{et al.} propose an adaptive Wi-Fi intrusion prevention system whose
context- and score-aware policy engine enforces real-time mitigation once an
attack is flagged \cite{n33}. That design is complementary to ours on the
mitigation side, but it runs as a fixed-location prevention layer and neither
partitions the detector nor preserves detection continuity as the endpoint roams
across domains. Lone \textit{et al.} survey machine-learning misbehavior
detection for connected and autonomous vehicles and explicitly identify
compromise during handover and roaming, as well as spatial partitioning of the
monitored region, as open issues \cite{n17}. MAPLE-Sec offers a concrete
mechanism for exactly that gap, migrating anchored inference state across edge
nodes so detection survives the handover that the survey flags as unresolved.

Across the three works, MAPLE-Sec is distinguished by jointly providing a
device-to-edge partition that adapts to mobility, state migration that preserves
detection continuity across handovers, and latency-bounded inference suitable for
constrained endpoints. The first two properties are absent from all three recent
approaches, and the combination is what allows MAPLE-Sec to sustain detection
under motion rather than treating mobility as an exception.

\section{Conclusion}\label{sec:conc}
In this paper, we have proposed MAPLE-Sec, a mobility-aware partitioned
language-model engine for real-time threat detection on roaming IoT endpoints,
directly addressing the handover-driven detection gap identified as a key open
challenge in mobile IoT security. We have observed that binding detection state
to a fixed location creates handover-induced blind spots, and that uncompressed
offload and fixed-depth inference waste the scarce uplink, energy, and latency
budgets of moving devices. To preserve detection continuity, we have designed an
activation-compression codec (ACC) that migrates inference state across edge
nodes; to control uplink and energy under motion, we have designed a
trajectory-conditioned partition scheduler (TPS); and to meet millisecond latency
budgets, we have designed a token-budgeted early-exit mechanism (TBEE). Extensive
experiments on CIC-IoT and Edge-IIoTset under realistic SUMO-generated
mobility have validated the effectiveness of MAPLE-Sec, demonstrating a 9.3\%
relative F1 improvement over Cloud-only and a 2.9 percentage-point gain over
EdgeShard, with $2.5\times$ lower latency. Ablation shows that the ACC codec
reduces migration energy by $2.4\times$ over raw activation transfer, with
near-linear scaling to dense edge deployments. A pivotal direction for future research involves jointly learning
the codec and the anchor representation end-to-end across mobility traces. In
parallel, we plan to extend MAPLE-Sec to multi-endpoint cooperative detection,
where neighboring roaming devices share migrated anchors to defend against
coordinated attacks. The broader vision is a fully distributed, mobility-resilient
intrusion detection layer for heterogeneous wireless IoT infrastructure, where
detection continuity is treated as a first-class design objective alongside
latency and energy, rather than an afterthought to be resolved by cloud fallback.

MAPLE-Sec has three principal limitations that future work should address. First,
the mobility model is constructed by replaying stationary traffic captures onto
SUMO-generated vehicular traces; flows are not natively generated on moving
devices, so the statistical interdependence between mobility state and traffic
pattern is approximated rather than measured. Second, although the device head is
executed on a physical Jetson Orin Nano and the link emulator uses calibrated
\texttt{tc netem} parameters, the edge tail and ACC decoder run on a desktop GPU;
end-to-end system evaluation on a fully embedded edge node (e.g., Jetson AGX
Orin) remains as future work. Third, the ACC codec was compared in the ablation
against raw activation migration but not against quantization-only baselines
(8-bit, 4-bit) or established learned image-compression codecs; this comparison
will be carried out in our future work. Addressing these limitations will
require natively mobile traffic generation, fully embedded deployment, and a
broader codec comparison study. As a potential future direction, we are looking forward to extending our method to improve the performance of various applications, such as large language
models~\cite{fang2026sibyl,lin2024split,fang2024automated,zhan2025prism} and distributed learning
system~\cite{lin2026hasfl,lyu2023optimal,fang2026nsc}.

\bibliographystyle{IEEEtran}
\bibliography{ref}

\end{document}